\documentclass[iop]{emulateapj}
\usepackage{natbib}
\usepackage{amsmath}
\usepackage{graphicx}
\usepackage{epstopdf}
\usepackage{upgreek}
\usepackage{multirow}
\usepackage{booktabs}
\usepackage{hyperref}
\usepackage{color}
\usepackage[figuresright]{rotating}
\usepackage{subfigure}

\graphicspath{ {../figures/save/} }
\newcommand{\HII}{\mbox{H\,\textsc{ii}}}%
\newcommand{\mt}{$\upmu$m~}
\newcommand{\m}{$\upmu$m}
\newcommand*{\thead}[1]{\multicolumn{1}{c}{#1}}
\slugcomment{Accepted, August 19 2015}

\shorttitle{Probing charge with the 15-20 \mt PAH bands}
\shortauthors{Shannon et al.}

\begin{document}

\title{Probing the ionization states of polycyclic aromatic hydrocarbons via the 15-20 \mt emission bands}

\author{M. J. Shannon\altaffilmark{1,2}, D. J. Stock\altaffilmark{2}, E. Peeters\altaffilmark{2,3}}
\altaffiltext{1}{Contact: mshann3@uwo.ca}
\altaffiltext{2}{Department of Physics and Astronomy, University of Western Ontario, London, ON, N6A 3K7, Canada}
\altaffiltext{3}{SETI Institute, 189 Bernardo Avenue, Suite 100, Mountain View, CA 94043, USA}

\begin{abstract}

We report new correlations between ratios of band intensities of the 15-20 \mt emission bands of polycyclic aromatic hydrocarbons (PAHs) in a sample of fifty-seven sources observed with Spitzer/IRS. This sample includes Large Magellanic Cloud point sources from the SAGE-Spec survey, nearby galaxies from the SINGS survey, two Galactic ISM cirrus sources and the spectral maps of the Galactic reflection nebulae NGC 2023 and NGC 7023. We find that the 16.4, 17.4 and 17.8 \mt band intensities are inter-correlated in all environments. In NGC 2023 and NGC 7023 these bands also correlate with the 11.0 and 12.7 \mt band intensities. The 15.8 \mt band correlates only with the 15-18 \mt plateau and the 11.2 \mt emission. We examine the spatial morphology of these bands and introduce radial cuts. We find that these bands can be spatially organized into three sets: the 12.7, 16.4 and 17.8 \mt bands; the 11.2, 15.8 \mt bands and the 15-18 \mt plateau; and the 11.0 and 17.4 \mt bands. We also find that the spatial distribution of the 12.7, 16.4 and 17.8 \mt bands can be reconstructed by averaging the spatial distributions of the cationic 11.0 \mt and neutral 11.2 \mt bands. We conclude that the 17.4 \mt band is dominated by cations, the 15.8 \mt band by neutral species, and the 12.7, 16.4 and 17.8 \mt bands by a combination of the two. These results highlight the importance of PAH ionization for spatially differentiating sub-populations by their 15-20 \mt emission variability.

\end{abstract}


\section{Introduction}

Polycyclic aromatic hydrocarbons (PAHs) are the most common polyatomic molecules in the universe \citep{tielens2013_moleuniverse} and are thought to produce prominent infrared emission bands at, e.g., 3.3, 6.2, 7.7, 8.6, 11.2 and 12.7 \m. These emission bands are seen in many different environments, including \HII~regions, reflection nebulae (RNe), planetary nebulae (PNe), and the diffuse interstellar medium (ISM). They are attributed to infrared (IR)-active vibrational modes, generally stretching or bending modes of C-C and/or C-H bonds. Bending modes of the carbon skeleton, denoted as C-C-C modes, produce weaker but identifiable emission features at 15.8, 16.4, 17.4, and 17.8 \m, typically perched atop a broad 17 \mt plateau (e.g.,  \citealt{moutou2000,vankerckhoven2000,vankerckhoven2002}).

PAH emission bands can be separated into three categories. The 3-15 \mt bands arise from nearest-neighbour vibrations, such as the 3.3 \mt C-H stretch or the 8.6 \mt in-plane C-H bend. These nearest-neighbour vibrations are common to all PAHs, and their collective emission features at similar wavelengths leads to blended emission. As a result, the emission intensities are high and they are readily observed in astronomical environments. In contrast, vibrations of the entire molecule (typically occuring at $> 20$\m, such as ``drumhead" modes) are extremely molecule-dependent (e.g., \citealt{vankerckhoven2000,boersma2010,ricca2010,boersma2011}). The intensities of these bands are very low and they are difficult to detect; in fact, none have, to date, been identified in space. The 15-20 \mt bands however represent an intermediate between the strong 3-15 \mt bands and the rich but weak 20+ \mt bands; they arise from vibrational modes involving a subset of the molecule (e.g., \citealt{vankerckhoven2000,peeters2004b,boersma2010}). They are generally thought to arise in PAHs with $\sim50-200$ carbon atoms \citep{boersma2010}. The plateaus upon which the PAH bands sit are thought to arise from emission by large PAHs or PAH clusters (e.g., \citealt{allamandola1989,tielens2008}).

The 3-15 \mt bands are known to have a series of correlations between the intensities of specific bands, typically attributed to common properties (primarily ionization state). The 15-20 \mt PAH bands however have thus far not exhibited any correlations amongst themselves (e.g., \citealt{smith2007,boersma2010,ricca2010,peeters2012}, hereafter referred to as paper \textsc{I}), though there are results showing the 16.4 \mt band correlates with the 6.2 and 12.7 \mt bands (\citealt{boersma2010}, paper \textsc{I}). Previous studies have had limited sample sizes of moderate quality or focused on single objects. Here we investigate correlations in a large sample of objects to examine the behaviour of the 15-20 \mt bands, and their relationship to the 3-15 \mt emission features. The paper is organized as follows: the observations and data reduction are presented in Section 2. The methods used to analyze the spectra are summarized in Section 3. Results, including correlation plots and spatial maps, are presented in Section 4. We interpret these results in Section 5 and summarize our findings in Section 6.

\section{Observations and Data Reduction}

\subsection{Target selection}
\label{sec:targets}
\begin{figure}
	\centering
	\includegraphics[clip=true, trim=5cm 0.7cm 6cm 0.5cm, width=1.0\linewidth]{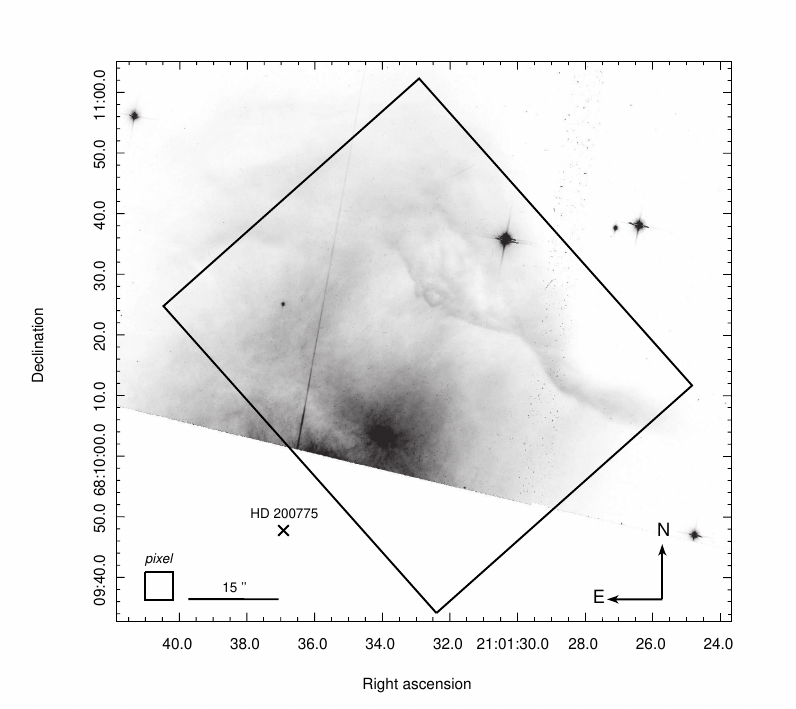}		
	\caption{
	Hubble Space Telescope image of NGC 7023, with the Spitzer/IRS SH field of view outlined by the black box. This image is composed from four filters: wide-band filters at 475 and 625 nm, a long-pass filter at 850 nm, and a narrow-band filter at 658 nm. The exciting star, HD 200775, a B2.5 Ve star \citep{finkenzeller1985}, is indicated by the black cross towards the lower left.
	\vspace{0.2cm}
	}
	\label{fig:ngc7023}
\end{figure}

We obtained spectra of a variety of objects from the SAGE-Spectroscopy (SAGE-Spec) Spitzer legacy program \citep{kemper2010} and sources from the Spitzer Infrared Nearby Galaxies Survey (SINGS, \citealt{kennicutt2003}). We chose sources displaying clear PAH emission in the 15-20 \mt region. Such emission can be identified by a wide 17 \mt complex (or plateau) and narrower features at 15.8, 16.4, 17.4 and 17.8 \m. Sources more extended than 13.5\arcsec~in the SAGE-Spec sample were excluded, as this corresponds to the approximate size of the SL aperture at 11.2 \mt\ (which is used for normalization when comparing band intensity ratios). Only sources with a 3$\sigma$ detection of their 15-20 \mt emission components (or a subset) are included. We also applied this measurement criterion to the ``MIR" (5-15 \m) PAH bands. Fifty-seven sources were identified as meeting these selection criteria, including two ISM cirrus sources. The sources in our sample are presented in Table~\ref{table:pointsources}. We also analyze a high-resolution spectral map of the RN NGC 7023 (Figure~\ref{fig:ngc7023}, previously analyzed by \citealt{rosenberg2011,berne2012,boersma2013,boersma2014,boersma2015}). We compare our results to that of NGC 2023 (paper \textsc{I} and Peeters et al. 2015, in prep., hereafter referred to as paper \textsc{II}). The 15-20 \mt measurements of \cite{boersma2010}, comprising 15 sources, are also included for comparison. Our total sample includes carbon-rich evolved stars, young stellar objects and \HII~regions in the Large Magellanic Cloud, nearby galaxies, Galactic RNe, and other ISM sources.

\subsection{Observations}
The spectra we analyze were acquired with the Infrared Spectrograph (IRS; \citealt{houck2004}) on board the Spitzer Space Telescope \citep{werner2004}. The IRS instrument provides high- and low-resolution spectroscopy in the mid-IR regime ($5-40$ $\upmu$m). Low-resolution spectra ($R \sim60-130$) were obtained with the Short-Low (SL) and Long-Low (LL) modules, which provide coverage from $5-14$ $\upmu$m and $15-40$ $\upmu$m, respectively. We also analyze a high-resolution ($R \sim600$) spectral map of NGC 7023 and 2 ISM objects acquired with the IRS/Short-High (SH) module, covering the wavelength range $9.9-19.5$ \m. These SH data were obtained from the NASA/IPAC Spitzer Heritage Archive\footnote{\url{http://sha.ipac.caltech.edu/applications/Spitzer/SHA/}} (AORkey: 3871232, PI: Giovanni Fazio). Previous studies that have examined these data are those of \cite{sellgren2007,rosenberg2011,boersma2013,boersma2014,boersma2015}.

\begin{table*}
	\centering
    \caption{Sample}
	\begin{tabular}{l c c c r c r r}
    \toprule
    \toprule
    Object & Type & Resolv.$^a$ & SSID$^b$ & AOR key & Extent$^c$ [$\arcsec$] & \multicolumn{1}{c}{RA [J2000]}  & \multicolumn{1}{c}{Dec. [J2000]} \\ 
	\midrule
    HD269211 	&	\HII?$^d$ 	&	Integ. & 4250	&	19151360	&	9.7	&	05 12 30.26   	& -70 24 21.75	\\
    HD32364 	&	\HII?$^d$ 	&	Integ. & 4122	&	19150592	&	11.2	&	04 57 14.28   	& -68 26 30.55	\\
    IRAS05192-6824 	&	 UC\HII $^e$ 	&	Integ. & 4313	&	19011584	&	6.1	&	05 19 06.84    	& -68 21 36.41	\\
    IRSX104 	&  \HII	 	& Integ.	 & 104 & 22425088	&	-	&	05 24 13.30 & -68 29 58.98	\\
    IRSX4087 	&	\HII 	&	Integ. & 4087	&	11239168	&	5.3	&	04 55 06.47   	& -69 17 08.30	\\
    IRSX4713 	&	 YSO$^f$ 	&	Integ. & 4713	&	16943616	&	6.3	&	05 40 12.02   	& -70 10 05.83	\\
    IRSX4729 	&	 YSO$^f$ 	&	Integ. & 4729	&	16943616	&	5.9	&	05 40 46.83   	& -70 11 22.73	\\
    IRSX4755 	&	 YSO$^f$ 	&	Integ. & 4755	&	11239424	&	7.0	&	05 43 19.62   	& -69 26 27.68	\\
    MSXLMC1306 	&	\HII 	&	Integ. & 4059	&	10958592	&	6.5	&	04 52 58.80   	& -68 02 56.77	\\
    MSXLMC217 	&	\HII 	&	Integ. & 4263	&	10962176	&	5.9	&	05 13 24.68    	& -69 10 48.12	\\
    MSXLMC559 	&	\HII 	&	Integ. & 4419	&	10963712	&	-   	&	05 25 49.23    	& -66 15 08.50	\\
    MSXLMC764 	&	\HII 	&	Integ. & 4538	&	10967040	&	7.6	&	05 32 52.62    	& -69 46 22.84	\\
    MSXLMC836 	&	\HII 	&	Integ. & 4527	&	10966528	&	6.4	&	05 32 31.95   	& -66 27 15.15	\\
    MSXLMC889 	&	\HII 	&	Integ. & 4621	&	10968576	&	6.7	&	05 38 31.64    	& -69 02 14.92	\\
    MSXLMC934 	&	\HII 	&	Integ. & 4652	&	10969088	&	13.4	&	05 39 15.83  	& -69 30 38.39	\\
    N159-P2 	&	 YSO 	&	Integ. & 4665	&	12548352	&	10.8	&	05 39 41.88   	& -69 46 11.94	\\

	NGC 7023 & RN & Extend. & - & 3871232 & - &	21 01 32.50 & +68 10 18.0  \\
	Cirrus1b & ISM & Extend. & - & 4120832 & - & 17 32 53.80  & -33 10 01.4  \\
	Cirrus3  & ISM & Extend. & - & 4119296 & - & 16 03 38.70  & -52 18 01.7  \\

    \bottomrule
    \end{tabular}
	\parbox{0.8\linewidth}{
	\vspace{0.1cm}
	\footnotesize \centering 
	\begin{flushleft}
	{\bf Notes.} Additional sources (not listed) are from \cite{smith2007,boersma2010,peeters2012}. $^a$Whether the source is resolved or not (i.e., integrated spectrum or extended); integrated sources are from Surveying the Agents of Galaxy Evolution Spectroscopy program (SAGE-Spec; \citealt{kemper2010}), data release 3. $^b$SAGE-Spec ID number. $^c$ Point sources are those with no extent listed. $^d$Object type uncertain in literature. $^e$Ultra-compact \HII~region \citep{beasley1996}. $^f$Classification from Woods et al. (in prep.).
	\end{flushleft}
	}
	\label{table:pointsources}
\end{table*}

\subsection{Reduction}
The SAGE-Spec data included in this study were obtained in reduced form. The project and the data reduction process are described in detail by \citet{kemper2010}; we briefly review the reduction steps here. The Spitzer Science Center\footnote{\url{http://ssc.spitzer.caltech.edu}} (SSC) data-reduction pipeline (version S18.7) was used to produce flat-field corrected images. Background subtraction was completed using on- and off-source nod positions. Rogue pixels were removed and replaced using the \texttt{imclean} algorithm, part of \texttt{irsclean}, provided by the SSC. Spectra were extracted using the SSC pipeline modules \texttt{profile}, \texttt{ridge}, and \texttt{extract}. Extracted spectra were coadded to produce one spectrum per nod position, per instrument module. The products were then calibrated and stitched together to produce the final spectra.

The SINGS data were also provided in reduced form by \cite{smith2007}. These data were processed with IRS pipeline version S14 and reduced with the \texttt{CUBISM} tool \citep{smith2007cubism}, including scaling and stitching of spectral segments to construct the final low-resolution spectra.

The SH maps of NGC 7023, Cirrus 1b and Cirrus 3 were reduced in a similar manner to that of NGC 2023, which was described in paper \textsc{I}. The raw data were processed by the SSC, using pipeline version S18.7. The reduction process was completed using \texttt{CUBISM} \citep{smith2007cubism}, including co-addition and bad pixel cleaning. The extraction for NGC 7023 was accomplished by stepping a $2\times 2$-pixel aperture across the map.

\section{Analysis}
\subsection{Spectral inventory}

\begin{figure*}
	\centering
	\includegraphics[width=0.8\linewidth]{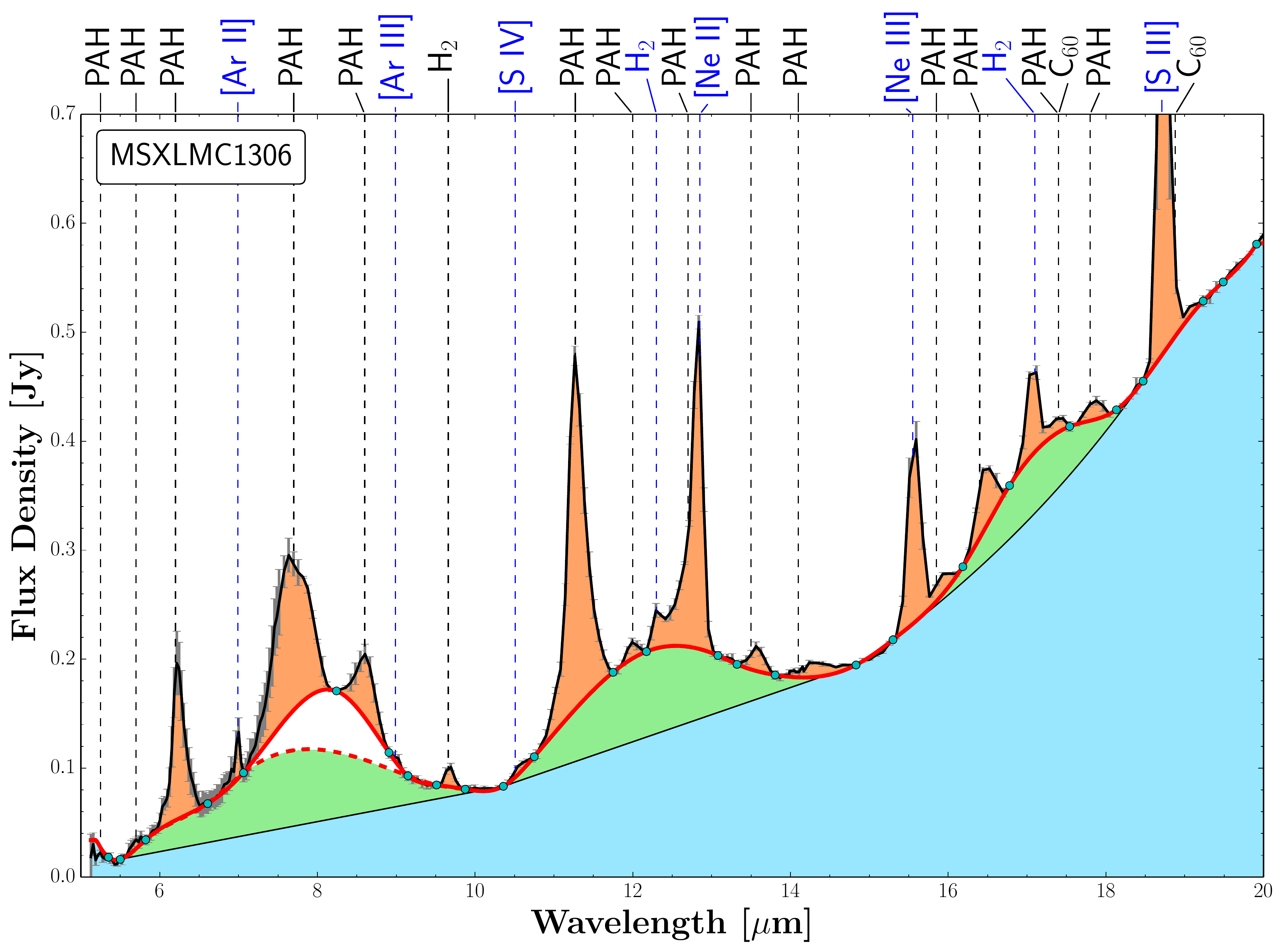}
	\caption{
	A typical IRS low-resolution spectrum from the SAGE-Spec sample is shown, that of LMC \HII~region MSX 1306. Prominent PAH features and atomic/molecular emission lines are visible (see labels at top). A local continuum is identified with a spline fit (red line). The continuum in the 5-10 \mt range can be fit with or without a continuum point between the 7.7 and 8.6 \mt features (see dashed red line).
	Three plateaus are identified underneath the PAH emission bands (green shaded region).
	}
	\label{fig:typical_spectrum}
\end{figure*}

A typical IRS low-resolution spectrum is shown in Figure~\ref{fig:typical_spectrum}, represented by the \HII~region MSX LMC 1306. A rising mid-to-far-IR dust continuum is common in many objects in the sample. There are also several atomic emission lines frequently present, including [Ar~\textsc{ii}] 6.99 \m, [Ar~\textsc{iii}] 8.99 \m, [S~\textsc{iv}] 10.51 \m, [Ne~\textsc{ii}] 12.81 \m, [Ne~\textsc{iii}] 15.56 \m, and [S~\textsc{iii}] 18.71 \m. We frequently observe the S(3) 9.7 \m, S(2) 12.3 \mt and S(1) 17.0 \mt rotational lines of H$_2$, and occasionally C$_{60}$ fullerene emission at 17.4 and 18.9 \mt \citep{cami2010,sellgren2010}. The PAH features at 6.2, 7.7, 8.6, 11.2, 12.7, 15.8, 16.4, 17.4, and 17.8 \mt are visible in all sources. Note that the 17.4 \mt band arises from both PAH and C$_{60}$ emission \citep{sellgren2010}. In addition to the aforementioned PAH emission bands, there are several weaker PAH bands visible at 5.25, 5.75, 6.0, 11.0, 12.0, 13.5 and 14.2 \m. Strong emission plateaus from 5-10 \m, 10-15 \m, and 15-18 \mt are observed. Additionally, some spectra display a strong silicon carbide feature at 11 \mt \citep{treffers1974,bernardsalas2009}.

\subsection{Spectral decomposition}

A local spline continuum is fit to each spectrum to isolate the PAH emission features (Figure~\ref{fig:typical_spectrum}), a method that has been used in several previous studies (e.g., \citealt{hony2001,peeters2002,bernardsalas2009,boersma2010,stock2013,stock2014}). Spline anchor points are chosen to be adjacent to emission features, and are allowed to move within a predefined window. When necessary, smoothing and/or a process that locates local minima are used to improve anchor positions. In addition to the local continuum, a global continuum is defined. This continuum underscores the emission plateaus at 5-10 \m, 10-15 \m, and 15-18 \m. The 7.7 and 8.6 \mt features can be drawn with or without a spline anchor point between them, depending on the decomposition. Here, we use such an anchor point near $8.2$ \m. The molecular hydrogen and atomic emission lines will be dealt with separately and are discussed in Section~\ref{sec:methods}.

Other methods for measuring the PAH bands are possible and are considered in the literature. These include fitting Drude profiles to the individual components (the PAHFIT tool; \citealt{smith2007}) or Lorentzian profiles \citep{boulanger1998,galliano2008b}. It has been shown that the decomposition method chosen affects the measured band intensities, but it does not affect the trends found in the correlations of band intensity ratios \citep{smith2007,galliano2008b}. The spline decomposition method is chosen here for comparison with previous 15-20 \mt PAH studies (paper \textsc{I}, \textsc{II}, \citealt{boersma2010}).

\subsection{Measurement methods}
\label{sec:methods}

\begin{figure} 
	\includegraphics[width=\linewidth]{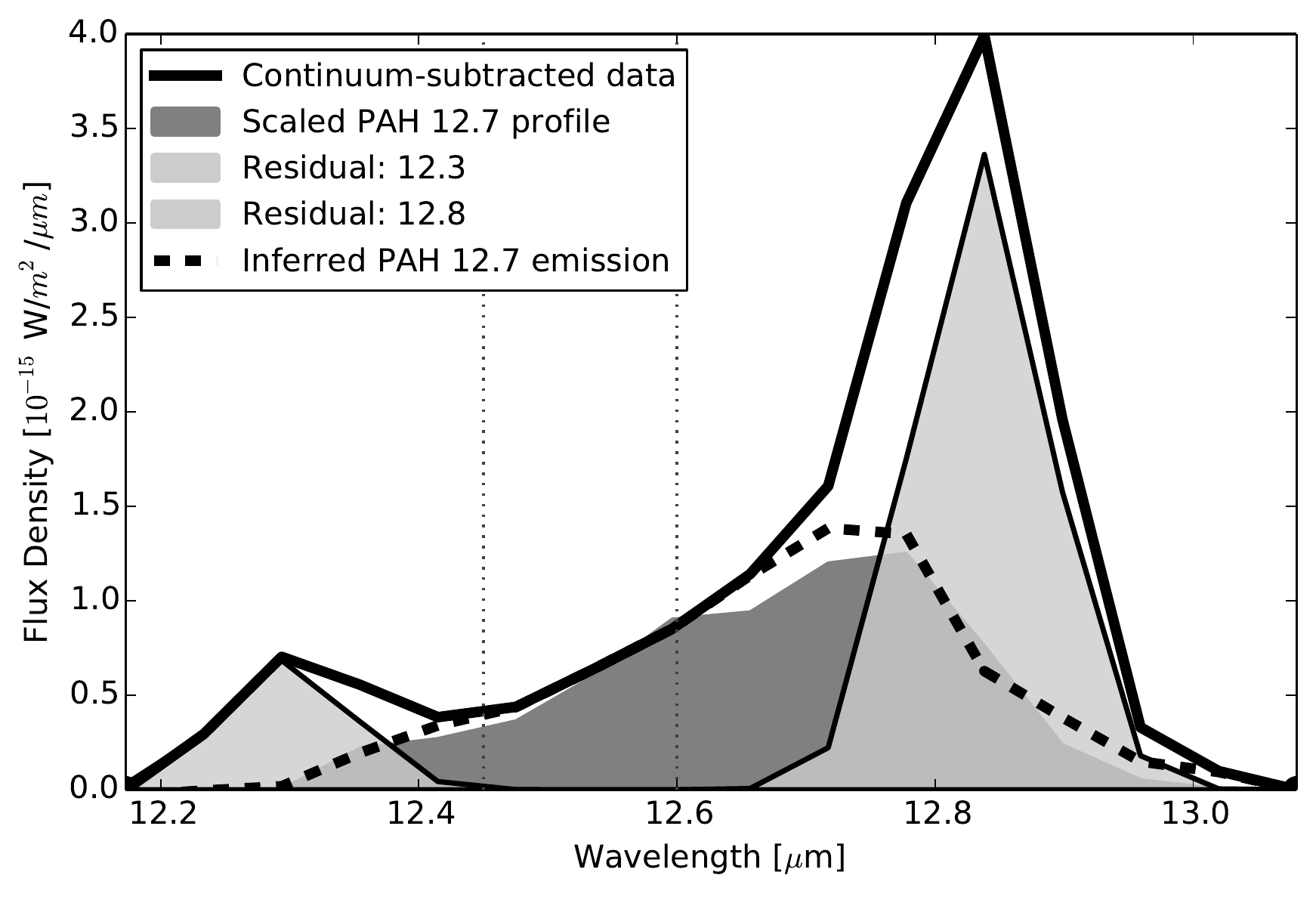}
	\caption{
	An example decomposition of the 12.7 \mt complex. The continuum-subtracted spectrum (solid line) is fit with the scaled mean 12.7 \mt PAH band profile (dark gray shading), as found by \cite{hony2001}. The fit is performed between 12.45 and 12.65 \mt (denoted by the dotted vertical lines). The fitted PAH profile (dark gray shading) can then be subtracted from the spectrum, leaving a remainder of the 12.3 \mt H$_2$ emission line and the 12.8 \mt [Ne \textsc{ii}] line (light gray shading). Returning to the original continuum-subtracted data, subtracting the two atomic lines then leaves only the inferred PAH 12.7 \mt emission (dashed line).
	}
	\label{fig:127_decomp}
\end{figure}

The plateaus are measured by direct integration between the global continuum and local continuum (see Figure~\ref{fig:typical_spectrum}). After subtracting the local continuum, PAH band fluxes in the 5-15 \mt range are measured by direct integration, except for the 6.0, 11.0 and 12.7 \mt bands. The 6.0 and 11.0 \mt bands are blended with the 6.2 and 11.2 \mt bands in low-resolution spectra, respectively. Therefore, Gaussians are used to separate these subfeatures and determine their fluxes (see paper \textsc{II} for details). Special care is required when measuring the 12.7 \mt PAH feature, as it blends with the 12.8 \mt [Ne \textsc{ii}] line and the 12.3 \mt H$_2$ line (when present). We separate these components by scaling the average 12.7 \mt PAH band profile (from \citealt{hony2001}) to fit the data between $12.45-12.6$ \m, where only PAH emission is expected (Figure~\ref{fig:127_decomp}). After subtracting the scaled PAH profile, the remaining lines are fit with Gaussians. This method is identical to the decomposition described by \cite{stock2014}, though we use the range $12.45-12.6$ \mt instead of $12.4-12.6$ \m. This is because the 12.3 \mt H$_2$ line is strong enough in some of our spectra to produce non-negligible (albeit minor) emission redward of 12.4 \m. We test for systematic effects in this decomposition by constructing an artificial spectrum. It is composed of the average 12.7 \mt PAH band profile from \cite{hony2001} plus two Gaussians, for the 12.3 H$_2$ line and the 12.8 [Ne \textsc{II}] line, and a variable amount of noise. We find that the derived PAH flux corresponds to the input artificial PAH flux within 2 to 3\% when the integrated strength of the [Ne \textsc{II}] line is less than ten times that of the 12.7 PAH band. This is true for all our sources. The 12.3 \mt H$_2$ line in our sample does not influence the derived PAH flux when decomposed this way.

The 15-20 \mt PAH bands and the atomic/molecular lines are also measured through Gaussian decomposition. The 16.4 \mt band is slightly asymmetric, while the other 15-20 \mt PAH bands are symmetric (paper \textsc{I}). There is frequent blending between the 15.55 \mt [Ne \textsc{iii}] line and the 15.8 \mt PAH feature, as well as between the 17.1 \mt H$_2$ line and the 17.4 \mt PAH band in low-resolution spectra. Emission by the fullerene C$_{60}$ is known to be present at 17.4 and 18.9 \mt in some objects \citep{cami2010,sellgren2010}. The 17.4 \mt C$_{60}$ emission is approximately half of the intensity of the 18.9 \mt band (estimated to be $\sim$0.4 in \citealt{sellgren2010}, $\sim$0.6 in \citealt{cami2010} and $\sim$0.5 in \citealt{bernardsalas2012}; we adopt 0.5). The 18.9 \mt band is observed in some of our spectra, typically blended to some degree with the 18.7 [S \textsc{iii}] line in low-resolution spectra. We examine the presence of C$_{60}$ emission by fitting the data with the model of C$_{60}$ emission from TC1 \citep{cami2010} and an appropriate Gaussian for the [S~\textsc{iii}] 18.71 \mt line. In the sample of integrated sources, we detect C$_{60}$ emission in four objects: IRSX 4755 (with associated signal-to-noise ratio, or SNR, of 3.4), IRSX 4087 (5.6), HD 32364 (5.5) and HD 269211 (5.9). We correct our measured 17.4 \mt band flux by subtracting the expected 17.4 \mt C$_{60}$ component when the 18.9 \mt C$_{60}$ band is present. The residual flux, if significant, is then attributed to the 17.4 \mt PAH band.

The measured PAH and C$_{60}$ fluxes of the SAGE-Spec sources and the two cirrus sources are presented in Table~\ref{table:fluxes_sage_pah}. The fluxes of the SINGS sources are shown in Table~\ref{table:fluxes_sings_pah}.

\section{Results}
\label{sec:results}

We examine correlations between the 15-20 \mt bands and the spatial distribution of the PAH emission in the extended RNe. NGC 2023S and NGC 2023N refer to the south and north maps of NGC 2023, respectively (papers \textsc{I} and \textsc{II}). Band fluxes in the correlation study are typically normalized to the 11.2 \mt band. Normalization is required because PAH abundances and object distances will vary from source to source. Only band ratios with 3$\sigma$ detections are included. The weighted Pearson correlation coefficients for all correlations are presented in Table~\ref{table:pearson_corrs}.

\begin{table*}
	\centering
	\caption{PAH emission band fluxes: SAGE and ISM sources}
	\resizebox{\linewidth}{!}{
	\begin{tabular}{l*{18}{rr}}
	\toprule
	\toprule
	\thead{Object} & \thead{6.2 \m} & \thead{7.7 \m} & \thead{8.6 \m} & \thead{11.2 \m} & \thead{12.7 \m} & \thead{15.8 \m} & \thead{16.4 \m} & \thead{17.4 \m} & \thead{17.8 \m} & \thead{Plateau} & \thead{18.9 \m~C$_{60}$} \\
	\midrule
HD269211 & - & - & 3.64 (1.14) & 13.09 (2.02) & 22.66 (1.51) & - & 1.77 (0.11) & 0.42 (0.11) & 0.56 (0.10) & 8.81 (0.88) & 1.07 (0.18) \\
HD32364 & - & - & - & 24.65 (1.68) & 23.01 (1.04) & - & 2.02 (0.12) & 0.42 (0.08) & 0.79 (0.14) & 16.91 (1.69) & 1.43 (0.26) \\
IRAS05192-6824 & 14.10 (0.77) & 28.35 (1.85) & 4.44 (0.98) & 14.75 (1.70) & 14.35 (1.56) & - & 1.54 (0.19) & - & - & 7.99 (0.80) & - \\
IRSX104 & 8.33 (1.53) & 10.40 (1.15) & - & 6.45 (0.71) & 3.28 (0.75) & 0.44 (0.12) & 0.72 (0.14) & - & - & - & - \\
IRSX4087 & 9.41 (1.15) & 14.48 (1.34) & 3.50 (0.75) & 10.60 (0.67) & 10.07 (0.74) & 0.63 (0.13) & 1.13 (0.15) & - & 0.48 (0.12) & 5.27 (0.53) & 1.62 (0.29) \\
IRSX4713 & 5.76 (0.64) & 8.47 (0.60) & 2.05 (0.20) & 5.43 (0.18) & 1.48 (0.18) & - & 0.41 (0.10) & - & - & - &  - \\
IRSX4729 & 6.95 (1.00) & 10.65 (1.23) & 2.17 (0.64) & 5.82 (0.57) & 1.92 (0.29) &-  & 0.41 (0.08) & - & - & 2.04 (0.20) & - \\
IRSX4755 & 18.30 (2.66) & 23.74 (3.48) & 5.13 (0.94) & 10.38 (1.17) & 6.02 (0.91) & - & 0.82 (0.11) & - & - & - & 0.34 (0.10) \\
MSXLMC1306 & 21.60 (5.09) & 38.72 (5.16) & 6.76 (1.38) & 20.59 (1.33) & 12.46 (0.86) & 0.48 (0.10) & 1.64 (0.10) & - & 0.46 (0.09) & 8.48 (0.85) & - \\
MSXLMC217 & 27.09 (4.94) & 52.02 (4.33) & 7.27 (2.00) & 27.14 (1.43) & 24.50 (1.25) & - & 2.30 (0.25) & - & - & 11.70 (1.17) & - \\
MSXLMC222 &-  &-  & 3.10 (0.33) & 6.97 (0.31) & - & - & 0.82 (0.11) & - &-  & - & - \\
MSXLMC559 & 31.00 (1.43) & 56.55 (1.39) & 7.92 (1.01) & 18.46 (0.87) & 17.97 (1.30) & - & 1.86 (0.61) & - & - & - & - \\
MSXLMC764 & - & 76.81 (13.82) & - & 39.66 (5.74) & 33.30 (3.61) & 1.85 (0.33) & 4.40 (0.18) & 0.96 (0.11) & 2.09 (0.31) & - & - \\
MSXLMC836 & 22.14 (0.67) & 34.98 (1.29) & 5.80 (0.52) & 19.88 (0.57) & 9.74 (0.41) & - & 1.28 (0.12) &  & 0.52 (0.13) & 7.88 (0.79) & - \\
MSXLMC889 & 61.10 (7.74) & 154.46 (6.41) & 20.51 (4.29) & 70.84 (5.38) & 80.30 (3.13) & - & 8.00 (0.17) & 1.47 (0.28) & 2.79 (0.34) & 36.50 (3.65) & - \\
MSXLMC934 & - & - & - & 50.93 (1.71) & 52.89 (2.34) & -  & 4.88 (0.54) & 0.97 (0.32) & 1.94 (0.47) & 23.38 (2.34) & - \\
N159-P2 & - & - & - & 32.70 (0.85) & 21.82 (0.99) & - & 5.01 (0.80) &  & 1.93 (0.63) & - & - \\
Cirrus1b & - & - & -  & 52.84 (1.08) & 33.84 (0.87) & 1.60 (0.29) & 3.60 (0.48) & 1.37 (0.28) & 1.30 (0.32) & - & - \\
Cirrus3 & - & - & - & 49.67 (0.94) & 27.65 (0.61) & 1.91 (0.18) & 3.38 (0.18) & 1.22 (0.23) & 0.95 (0.17) & - & - \\
	\bottomrule
	\end{tabular}
	}
	\\[0.9ex]
	\parbox{0.9\linewidth}{
	\footnotesize \centering
	\begin{flushleft}
	{\bf Notes.} In units of $10^{-16}$ W/m$^2$, except the two extended cirrus sources, which are in units of $10^{-2}$ W/m$^2$/sr. Uncertainties are given in parentheses.
	\end{flushleft}	
	}
	\label{table:fluxes_sage_pah}
\end{table*}

\begin{table*}
	\centering
	\caption{PAH emission band fluxes: SINGS sources}
	\resizebox{\linewidth}{!}{
	\begin{tabular}{l*{18}{r}}
	\toprule
	\toprule
	\thead{Object} & \thead{6.2 \m} & \thead{7.7 \m} & \thead{8.6 \m} & \thead{11.2 \m} & \thead{12.7 \m} & \thead{15.8 \m} & \thead{16.4 \m} & \thead{17.4 \m} & \thead{17.8 \m} & \thead{Plateau} \\
	\midrule
MRK33 & 7.82 (1.26) & 13.88 (1.49) & 3.33 (0.39) & 5.43 (0.30) & 3.72 (0.29) & - & 0.45 (0.09) & 0.16 (0.04) & 0.16 (0.05) & - \\
NGC0337 & - & 17.62 (1.62) & 3.58 (0.45) & 7.84 (0.38) & 3.83 (0.30) & - & 0.56 (0.10) & - & - & 3.47 (0.35) \\
NGC0628 & - & - & - & 3.48 (0.51) & - &-  & 0.25 (0.05) &-  & - & - \\
NGC0855 & - & - & - & 3.50 (0.90) & - & - & 0.31 (0.08) &-  & - &  -\\
NGC0925 & - & - & - & 4.72 (0.92) & - & - & 0.27 (0.06) &-  & - &  -\\
NGC1097 & 73.72 (1.36) & 157.03 (1.58) & 24.29 (0.47) & 65.07 (0.42) & 38.90 (0.38) & 1.07 (0.22) & 3.58 (0.42) & 1.56 (0.45) &-  & 25.74 (2.57) \\
NGC1482 & 100.70 (1.51) & 230.20 (2.04) & 31.10 (0.76) & 74.11 (0.71) &-  &  & 4.36 (1.29) &-  & - &  -\\
NGC1512 & - & - & - & 9.05 (1.48) &-  & - & 0.47 (0.08) &-  &-  & 3.42 (0.34) \\
NGC2403 & 7.51 (1.29) & 10.42 (1.37) & - & 5.13 (0.62) & - & - & 0.43 (0.05) & - & - & - \\
NGC2798 & - & - &-  & 47.55 (0.90) & 29.26 (0.84) & 0.83 (0.23) & 2.07 (0.35) & 1.10 (0.30) & - &  -\\
NGC2976 &-  &-  &-  & 4.56 (0.64) &-  & - & 0.38 (0.05) &-  & - & 1.83 (0.18) \\
NGC3049 &-  & - & - & 5.41 (0.75) & 3.25 (0.68) &-  & 0.28 (0.06) &-  & - & - \\
NGC3184 & - & - & - & 8.38 (1.08) & 5.47 (1.02) &-  & 0.31 (0.04) &-  & - & - \\
NGC3265 & - &-  &-  & 8.15 (0.75) & 3.89 (0.74) &-  & 0.46 (0.11) &-  & - &  -\\
NGC3351 & 16.01 (1.94) & 53.97 (2.09) & 5.30 (0.82) & 17.63 (0.84) & 11.64 (0.85) &  & 1.04 (0.15) & - & - & 12.92 (1.29) \\
NGC3521 & 13.44 (1.50) & 43.05 (1.72) & 6.01 (0.59) & 16.56 (0.52) & - & 0.33 (0.07) & 1.21 (0.10) & 0.36 (0.08) & 0.33 (0.10) & 13.29 (1.33) \\
NGC3621 & 15.28 (1.58) &-  & - & 12.12 (0.63) & 4.96 (0.67) &-  & 0.74 (0.06) & 0.21 (0.06) & - & 4.64 (0.46) \\
NGC3627 & 16.27 (1.86) & 27.32 (2.04) & 4.40 (0.71) & 19.59 (0.86) & - & 0.44 (0.08) & 0.86 (0.12) & - & - & 10.66 (1.07) \\
NGC3773 & - &-  & - & 4.12 (0.89) & - & - & - &-  &-  &  -\\
NGC3938 & - &-  & - & 3.00 (0.20) & 1.48 (0.20) & - & 0.18 (0.05) & - &-  & 2.18 (0.22) \\
NGC4254 & - & - & - & 23.23 (1.35) & - & - & 1.61 (0.16) & - & - & 8.40 (0.84) \\
NGC4321 & 33.00 (2.12) & 69.04 (2.58) & 10.03 (1.01) & 27.13 (1.07) & 14.83 (1.10) & - & 1.26 (0.15) & 0.54 (0.14) & - & 11.56 (1.16) \\
NGC4536 & 91.38 (2.19) & 192.04 (2.54) & 27.99 (0.79) & 66.01 (0.70) & - & 1.13 (0.36) & 2.86 (0.55) & - & - & 29.12 (2.91) \\
NGC4559 &-  & - & - & 5.29 (0.47) & 2.84 (0.41) & - & 0.36 (0.06) & 0.12 (0.04) & - &  -\\
NGC4569 & - & - & 5.89 (1.16) & 24.68 (1.24) & - & 0.44 (0.09) & 1.23 (0.09) &-  &-  & 14.15 (1.41) \\
NGC4579 & - & - & - & 4.76 (1.12) & - & - & 0.40 (0.11) & - & - &  -\\
NGC4631 & - & - & - & 30.72 (0.55) & - & 0.47 (0.13) & 2.09 (0.20) & - & - & - \\
NGC4736 & 34.32 (1.62) & 40.69 (2.13) & 9.73 (0.99) & 48.69 (1.08) & - & 0.89 (0.12) & 2.03 (0.18) & 0.47 (0.14) & 0.51 (0.12) & 16.99 (1.70) \\
NGC4826 & 44.40 (2.21) & 108.11 (2.58) & 18.10 (1.32) & 42.38 (1.44) & - & 0.93 (0.11) & 2.30 (0.18) & 0.72 (0.16) & 0.72 (0.10) & 19.41 (1.94) \\
NGC5033 & 19.75 (1.76) & 36.88 (1.75) & 6.15 (0.79) & 17.27 (0.81) & 8.47 (0.74) & - & 0.91 (0.14) & 0.33 (0.10) & 0.29 (0.09) & 9.01 (0.90) \\
NGC5055 & 16.65 (1.42) & 44.23 (1.76) & - & 21.37 (0.81) &-  &-  & 0.97 (0.10) & 0.30 (0.08) & 0.30 (0.06) & 10.06 (1.01) \\
NGC5194 & 27.32 (1.27) & 56.63 (1.59) & 10.35 (0.89) & 29.35 (0.88) & - & 0.48 (0.10) & 1.28 (0.14) & 0.43 (0.11) & 0.34 (0.09) & 12.83 (1.28) \\
NGC5195 & - & 32.48 (2.59) & - & 36.03 (1.28) & - & - & 1.93 (0.46) &-  & - & 15.19 (1.52) \\
NGC5713 & 36.36 (1.25) & 75.25 (1.62) & 11.23 (0.72) & 29.04 (0.69) & - & - & 1.78 (0.16) & - & - & 12.49 (1.25) \\
NGC5866 & - & - & - & 8.24 (0.86) & - & 0.22 (0.05) & 0.34 (0.07) & - & - &  -\\
NGC6946 & 90.92 (1.93) & 189.33 (2.10) & 29.22 (1.11) & 61.51 (1.05) & - & - & 4.34 (0.41) & - & - & 33.86 (3.39) \\
NGC7331 &-  &-  & - & 15.56 (1.26) & - & 0.27 (0.08) & 0.83 (0.08) & - & - & 7.93 (0.79) \\
NGC7552 & 81.57 (1.84) & 173.99 (2.22) & 26.90 (0.97) & 71.78 (0.96) & 47.74 (0.95) & - & 4.79 (0.97) & - & - &  -\\
	\bottomrule
	\end{tabular}
	}
	\\[0.9ex]
	\parbox{0.9\linewidth}{\footnotesize \centering
	\begin{flushleft}
	{\bf Notes.} In units of $10^{-14}$ W/m$^2$. Uncertainties are given in parentheses. No 3$\sigma$ detections of 18.9 \mt C$_{60}$ emission were found for these data.
	\end{flushleft}
}
	\label{table:fluxes_sings_pah}
\end{table*}

\begin{table}
	\centering
	\caption{Weighted Pearson correlation coefficients}
	\resizebox{\linewidth}{!}{
	\begin{tabular}{lllcc}
    \toprule
    \toprule
	Correlation 		& Integrated & N7023$^a$ 		& N2023N$^b$ & N2023S$^b$ \\
    &	 sources & & & \\
    \midrule
	17.4 vs. 16.4 	& 0.67			& 0.80			& 0.57	& 0.45 \\
	17.8 vs. 16.4 	& 0.93			& 0.85			& 0.72	& 0.82 \\
	17.8 vs. 17.4 	& 0.63			& 0.53			& 0.97	& 0.47 \\
    \midrule
	16.4 vs. 15.8 	& 0.74			& 0.22 			& -   	& -0.28 \\
	17.4 vs. 15.8 	& 0.58			& 0.37 			& -		& -0.14 \\
	17.8 vs. 15.8 	& 0.75			& 0.40 			& -		& -0.08 \\
	15.8 vs. 11.2	& -0.28			& 0.38			& -		& 0.89 \\
	Plat. vs. 11.2 & 0.42			& 0.47			& 0.92	& 0.87 \\
	Plat. vs. 15.8 & 0.28			& 0.46			& - 	 	& 0.88 \\
    \midrule
	12.7 vs. 11.0	& 0.51			& 0.95			& 0.88	& 0.91 \\
	16.4 vs. 11.0	& 0.36			& 0.92			& 0.92	& 0.95 \\
	17.4 vs. 11.0	& -0.14			& 0.84			& 0.69	& 0.51 \\
	17.8 vs. 11.0	& -0.26			& 0.87			& 0.89	& 0.81 \\
	16.4 vs. 12.7	& 0.28 [0.68]$^c$   & 0.89			& 0.88	& 0.92 \\
	17.4 vs. 12.7	& 0.69			& 0.83			& 0.55		& 0.35 \\
	17.8 vs. 12.7	& 0.23 [0.66]$^c$	& 0.86			& 0.53	& 0.71 \\
    \bottomrule
	\end{tabular}
    }
	\\[0.9ex]
	\parbox{1.0\linewidth}{\footnotesize \centering
	\begin{flushleft}
	{\bf Notes.} $^a$ See individual correlation plots for coefficients organized by region; regions defined in Figure~\ref{fig:ngc7023maps}. $^b$ See also paper \textsc{I}, \textsc{II}. $^c$ \HII~regions alone.
	\end{flushleft}	
	}
	\label{table:pearson_corrs}
\end{table}

\subsection{Correlation plots}

\subsubsection{The 16.4, 17.4 and 17.8 \mt bands}
\label{subsec:new_corrs}

\begin{figure}
	\includegraphics[width=\linewidth]{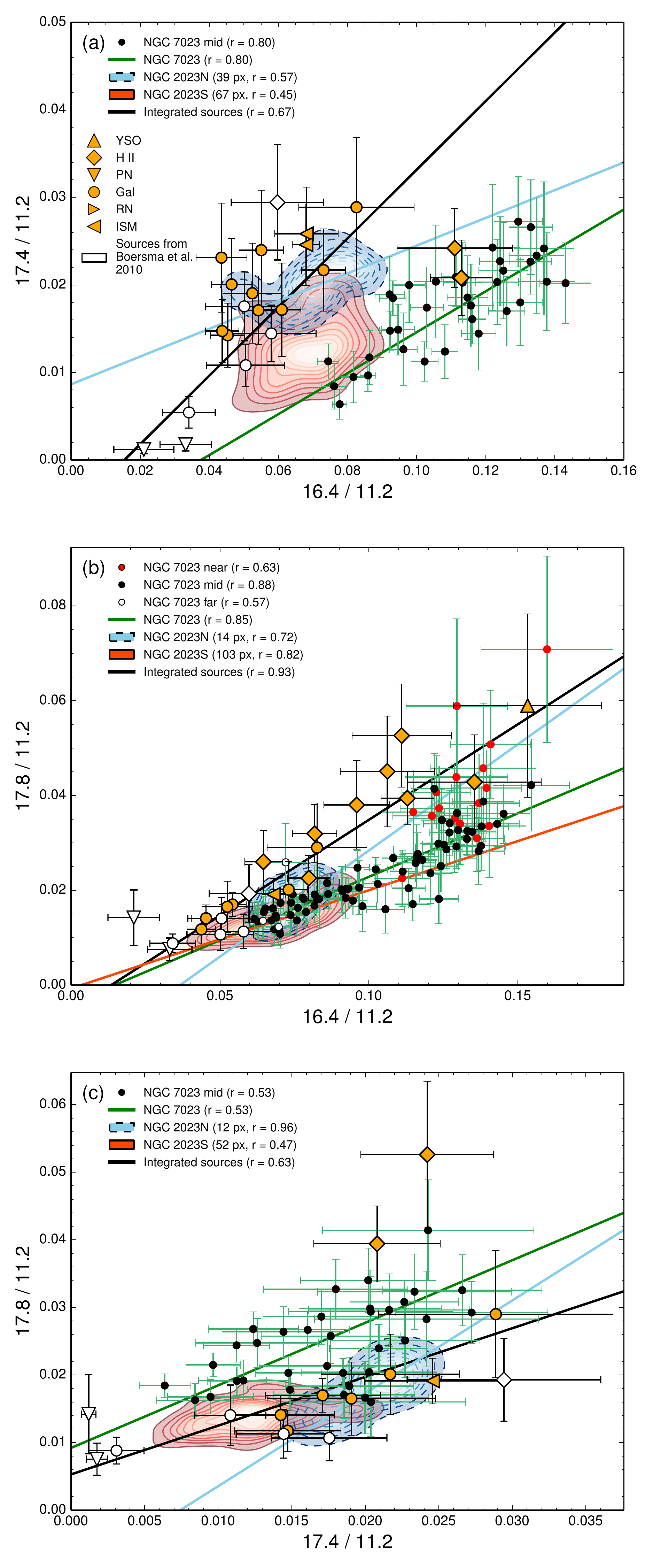}
	\caption{
	Correlation plots of flux ratios amongst the 16.4, 17.4 and 17.8 \mt features. The filled blue and orange density contours represent the number densities of measurements of NGC 2023 North and South, respectively (papers \textsc{I} and \textsc{II}). Each displays ten linear contour levels, computed with a Gaussian kernel. The data points with green errorbars are from NGC 7023, separated into three regions (defined in Figure~\ref{fig:ngc7023maps}): data nearest to the exciting star (red dots);	data from the center of the map, the peak PAH emission region (black dots); and data furthest from the exciting star (white dots). Each grouping is shown with an associated line of best fit when the correlation coefficient exceeds $r=0.50$: integrated sources (black line), NGC 7023 (green line), NGC 2023N (blue line), NGC 2023S (orange line).
	}
	\label{fig:all_corrs1a}
\end{figure}

Correlation plots of the 16.4, 17.4, and 17.8 \mt bands are presented in Figure~\ref{fig:all_corrs1a}. We address these in turn.

\textit{17.8 vs. 16.4 \m.} These band fluxes are the strongest correlated for the integrated sources (correlation coefficient $r=0.93$). Some segregation by object type is observed: two evolved stars sit at the lowest ratios of 16.4/11.2 and 17.8/11.2 (0.030 and 0.010, respectively), galaxies slightly higher at (0.060, 0.015), and \HII~regions at the highest values (centered near 0.10, 0.035). The galaxies and \HII~regions each span a range covering roughly a factor of two on each axis.
The 17.8 and 16.4 \mt bands also correlate in the RNe. NGC 7023 has a correlation coefficient of $r=0.85$ and spans nearly the same variation in 16.4/11.2 and 17.8/11.2 as the integrated sources. NGC 7023 has a cluster of data points centered near (0.13, 0.030), which we determine to originate in the portion of the map closest to the exciting star (the upper third of the map; see Figure~\ref{fig:ngc7023maps}). NGC 2023S has a correlation coefficient of $r=0.82$ and has a line of best fit close to that of NGC 7023, despite spanning half the range in abscissa; NGC 2023N has a similar range and correlation coefficient ($r=0.72$). Its line of best fit is closer to the fit for the integrated sources, but it should be noted it also has far fewer pixels than those contained in the NGC 7023 or NGC 2023S maps given the SNR criterion.

\textit{17.4 vs. 16.4 \m.} The 17.4 and 16.4 \mt band fluxes are observed to correlate. NGC 7023 exhibits a correlation coefficient of $r=0.80$, while NGC 2023S and 2023N have associated coefficients of $r=0.45$ and $r=0.57$, respectively. NGC 7023 and NGC 2023 have similar lines of best fit. The integrated sources however display a trend on a much steeper gradient, aside from two lone \HII~regions, and have a correlation coefficient of $r=0.67$. The integrated sources and NGC 2023 generally span the same abscissa and ordinate ranges, but NGC 7023 is set to higher ratios of 16.4/11.2, with only modest overlap with NGC 2023 and the integrated sources. This correlation was previously observed by \cite{berne2012}.

\textit{17.8 vs. 17.4 \m.} The 17.8 and 17.4 \mt bands also correlate. The integrated sources and NGC 7023 span similar ranges in abscissa and ordinate values, with associated coefficients $r=0.63$ and $r=0.53$, respectively. NGC 2023 spans half the range of 17.8/11.2 in comparison, though a similar extent is observed in 17.4/11.2. NGC 2023S has correlation coefficient of $r=0.47$ and NGC 2023N of $r=0.97$, though the latter has far fewer data points included (12 versus 52, respectively). The individual sources again fall upon a slightly different gradient, as they appear to be displaced toward higher 17.4/11.2 values. The outlying \HII~regions in this figure are also the outliers in the correlation plot of the 17.4 and 16.4 \mt bands. Systematic problems may affect the 17.4 \mt band measurement in our sample as its derived flux is dependent on an accurate measurement of the 18.9 \mt C$_{60}$ band. When the latter cannot be observed (or reliably measured) it is possible that the derived 17.4 \mt PAH band flux is overestimated, as the 17.4 \mt C$_{60}$ contribution cannot be removed, if present. This issue is particularly prominent in low-resolution spectra, which comprise most of the integrated sources in this sample (apart from the two ISM cirrus sources), as the 18.9 \mt C$_{60}$ band blends with the [S~\textsc{iii}] 18.71 \mt line. The RNe were observed in high resolution and are less susceptible to this type of overestimation, which may explain why the gradients observed in the correlation plots involving the 17.4 \mt band differ for integrated sources versus RNe (e.g., Figure~\ref{fig:all_corrs1a}).

\begin{figure}
	\includegraphics[width=\linewidth]{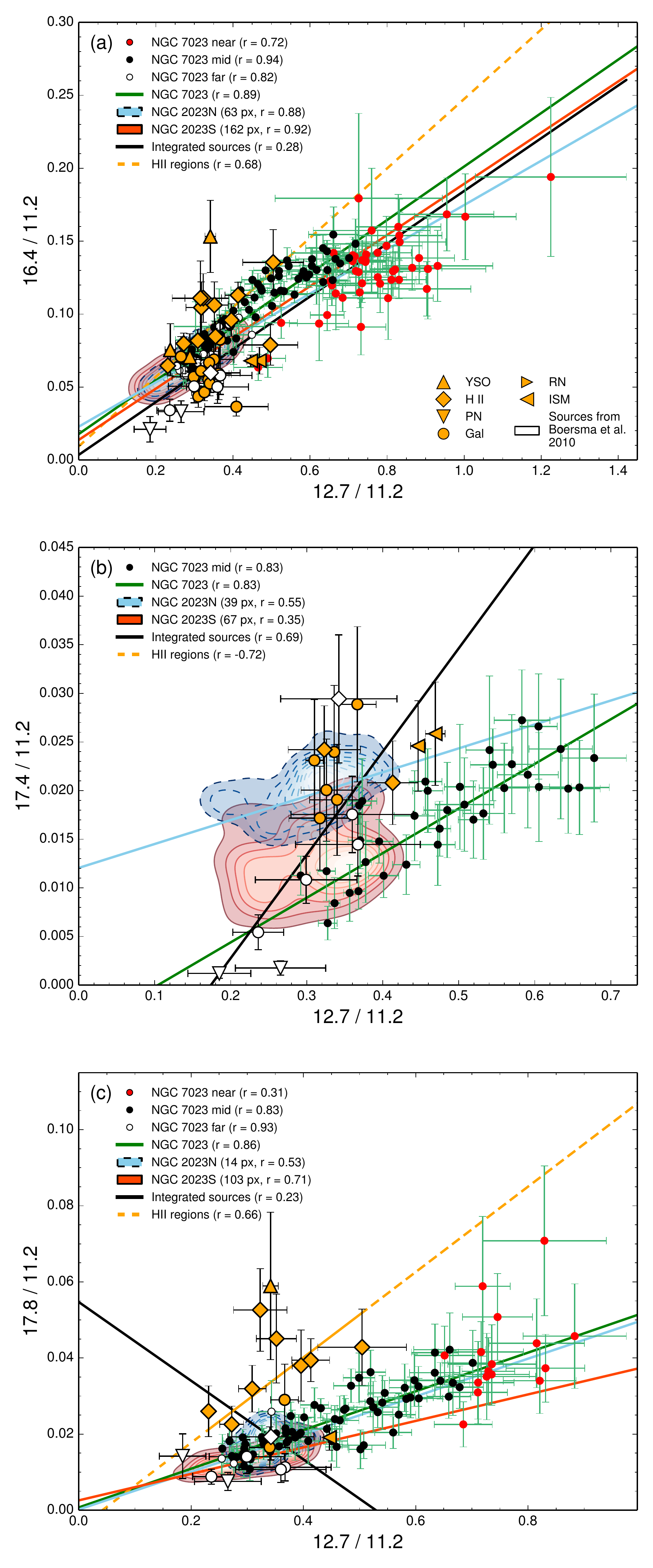}
	\caption{
	Correlation plots of the 16.4, 17.4 and 17.8 \mt band fluxes versus the 12.7 \mt band flux, each normalized to the 11.2 \mt band flux.
	}
	\label{fig:all_corrs2_1}
\end{figure}

\textit{Correlations with the 12.7 \mt band.} The 16.4, 17.4 and 17.8 \mt bands are observed to correlate with the 12.7 \mt band (Figure~\ref{fig:all_corrs2_1}). NGC 7023 displays a high correlation coefficient ($r=0.89$), though there is deviation from the linear fit in our data at high abscissa values. In our data, there is a cluster of data points at 12.7/11.2=0.75, 16.4/11.2=0.13, all of which originate in the third of the map nearest the exciting star in NGC 7023. Regardless, we find a strong correlation between these band ratios in NGC 7023, peaking in the middle portion of the map ($r=0.94$; see Figure~\ref{fig:ngc7023maps} for map region definitions). The gradient of the NGC 7023 data is consistent with those of NGC 2023 (both north and south). The integrated sources of this study show no trend as a whole, but they do tend to cluster about a position coincident with the center of the NGC 2023 data. The subset of \HII~regions alone however do display a correlation ($r=0.68$). Since the 16.4, 17.4 and 17.8 \mt bands correlate (Figure~\ref{fig:all_corrs1a}), it is not surprising that we find that the 17.4 and 17.8 \mt bands correlate with the 12.7 \mt band in RNe. We additionally observe that \HII~regions display a correlation between the 17.8 and 12.7 \mt bands. Nothing can be determined about \HII~regions amongst the 17.4 and 12.7 \mt bands as there are too few sources meeting the detection criteria. The 12.7 \mt band was previously observed to correlate with the 16.4 \mt band in NGC 2023 (paper \textsc{I}; $r=0.92$ in the southern map, $r=0.88$ in the northern map) and in NGC 7023 (\citealt{boersma2014}; their Figure 9).

\textit{Correlation with the 6.2 \mt band.} We examine the flux ratios of the 16.4 and 6.2 \mt bands in Figure~\ref{fig:hony} for the integrated sources in our sample. The 6.2 \mt band cannot be measured in the SH RNe data as the wavelength coverage begins at $\sim10$ \m.  Data from \cite{boersma2010} are included, where we have replaced the authors' average SINGS galaxy data point with our individual SINGS sources. We observe no linear correlation (r$=0.11$), though we note the lower-right quadrant is devoid of sources. Furthermore, we identify the \HII~regions as occupying the upper-left quadrant of this figure. If these are excluded from the data set, a modest correlation is observed (r$=0.47$). A mostly linear trend (r$^2=0.76$) between these bands has been identified in NGC 7023 by \cite{boersma2014}, their Figure 10, albeit with some residual curvature. Our measurements are consistent with their measurements at low abscissa/ordinate values.

Since we detect a correlation between the 16.4 and 12.7 \mt bands, and essentially none between the 16.4 and 6.2 \mt bands, we should expect no correlation between the 6.2 and 12.7 \mt bands for consistency. We indeed detect no such correlation in our data alone. However, the 6.2 and 12.7 \mt PAH bands have been observed to correlate for a sample of integrated sources of various object types \citep{hony2001} and within the RNe NGC 7023 \citep{boersma2014} and NGC 2023 (paper \textsc{II}). The lack of such a correlation in our sample of integrated sources is likely due to the small range of ratios probed by our data, as they span a small range of ratios relative to \cite{hony2001}.

\textit{Correlations with the 11.0 \mt band.} We observe that the 16.4, 17.4 and 17.8 \mt bands correlate with the 11.0 \mt band in NGC 2023 and NGC 7023 (Figure~\ref{fig:corrs_110}). The NGC 7023 pixels furthest from and nearest to the star tend to individually cluster, while the region containing the peak of the PAH emission shows linear correlations between band intensity ratios (regions defined in Figure~\ref{fig:ngc7023maps}). {This clustering has been previously observed, for example in \cite{boersma2014}. The 12.7 \mt band is also included for comparison, and the RNe display strong correlations between the 12.7 and 11.0 \mt bands ($r\geq0.90$). The integrated sources show no clear correlations between the 11.0 \mt band and the 12.7, 16.4, 17.4 and 17.8 \mt bands, as they tend to cluster near 11.0/11.2$\sim$0.060. They tend to be consistent with the observed RNe correlations for the 12.7 and 16.4 versus 11.0 \mt bands, but they deviate for the 17.8 \mt and especially the 17.4 \mt band.

\begin{figure*}
	\includegraphics[width=\linewidth]{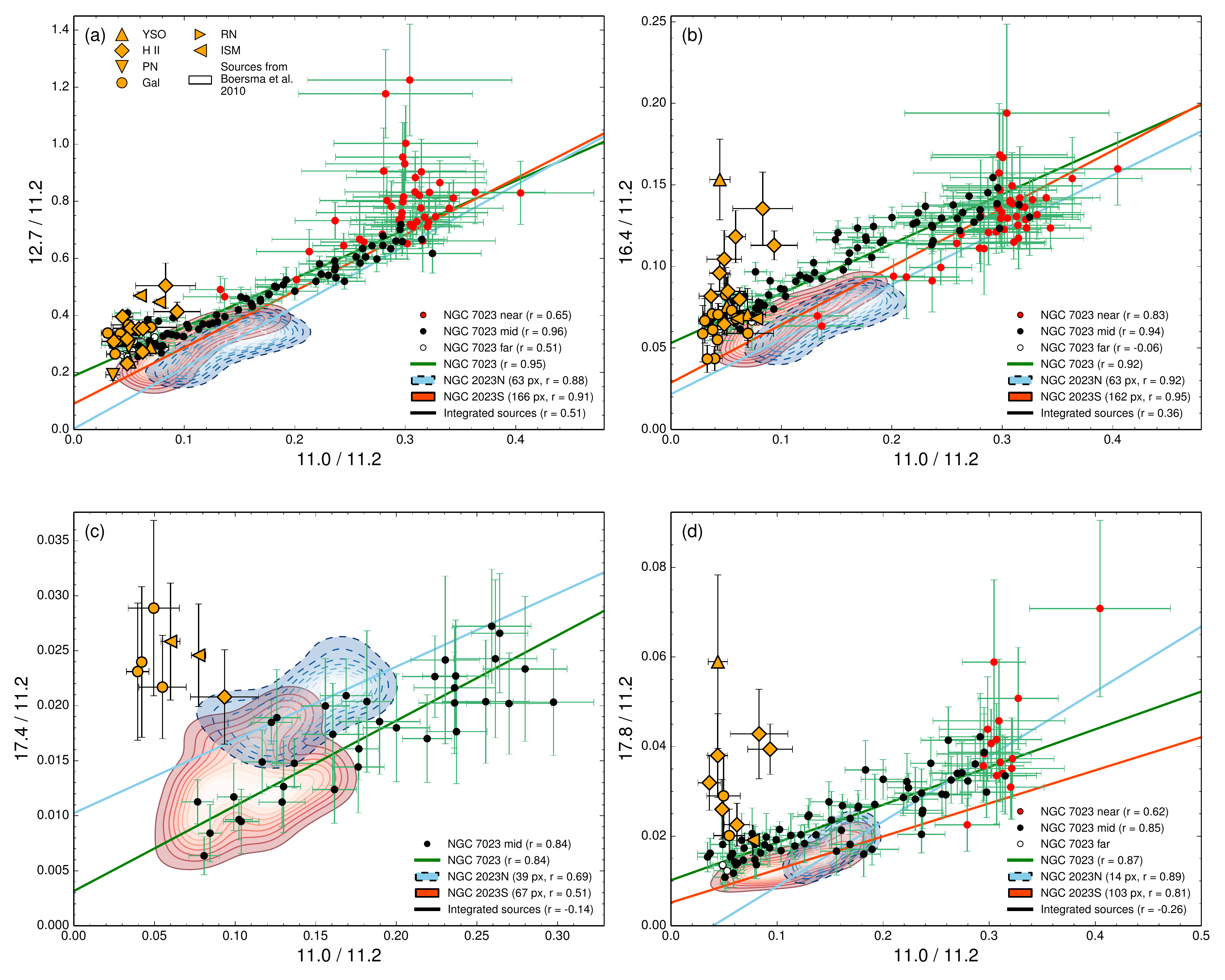}
	\caption{
	Correlation plots of the 12.7, 16.4, 17.4 and 17.8 \mt bands versus the 11.0 \mt band. 
	}
	\label{fig:corrs_110}
\end{figure*}

\begin{figure}
	\includegraphics[width=\linewidth]{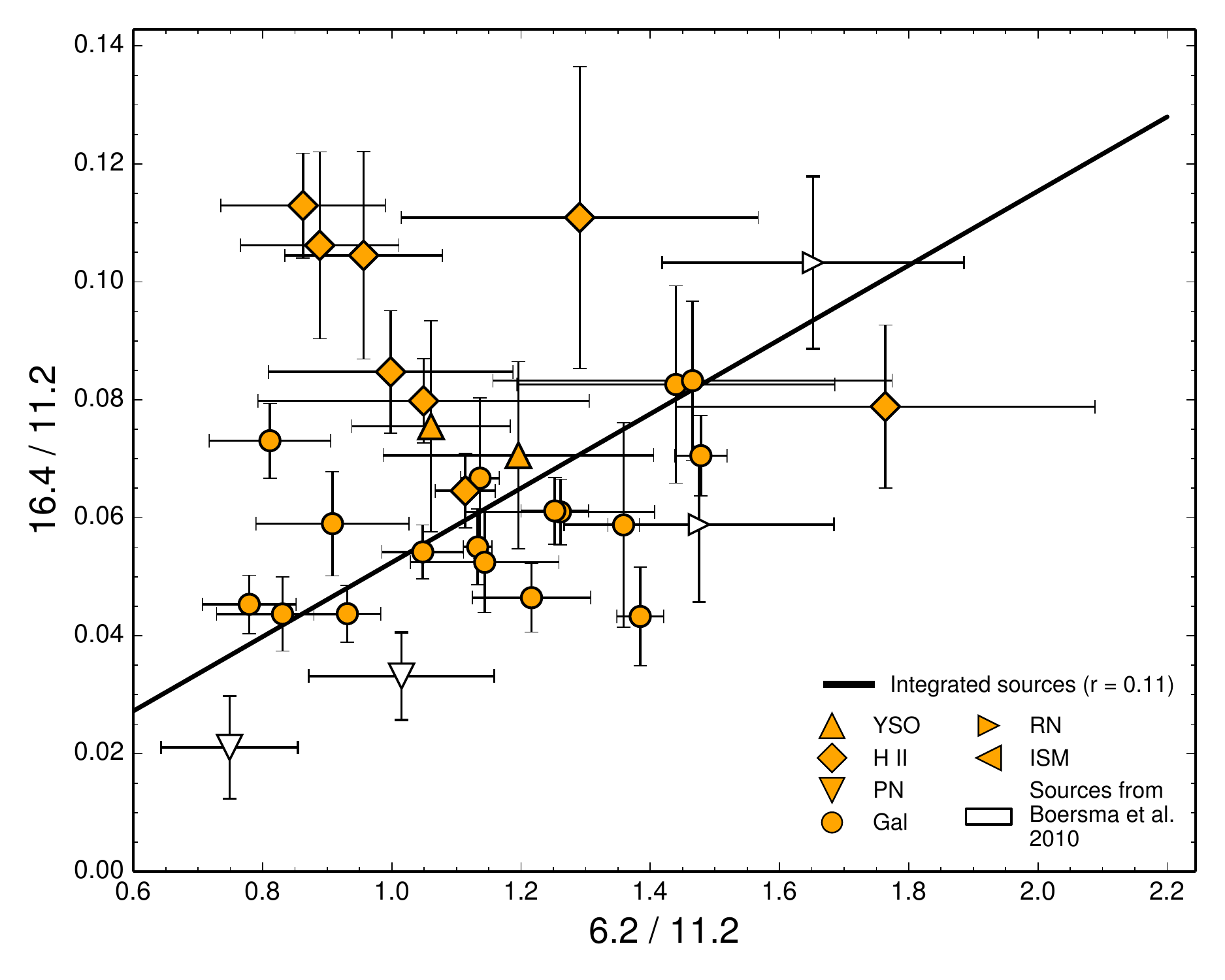}
	\caption{
	No linear relationships is observed between the 16.4 and 6.2 \mt bands ($r=0.11$); however, the lower-right quadrant is conspicuously devoid of sources, and our data span a small range of values (c.f. \citealt{boersma2014}).
	}
	\label{fig:hony}
\end{figure}

\subsubsection{The 15.8 \mt band}
\label{subsec:uncorrs}

The 16.4, 17.4 and 17.8 \mt bands do not appear to correlate with the 15.8 \mt band (Figure~\ref{fig:all_corrs1b}). NGC 2023S shows no correlation between these bands (e.g., $r=-0.28$ when comparing the 15.8 and 16.4 \mt bands) and NGC 2023N has too few detections to be represented here (papers \textsc{I} and \textsc{II}). NGC 7023 generally appears to have a similar distribution to NGC 2023S, with no overall correlation (e.g., $r=0.22$). However, NGC 7023 does seem to exhibit a ``C"-shaped distribution in the 16.4 versus 15.8 \mt band intensity correlation, and to a lesser degree in the 17.8 versus 15.8 \mt correlation plot. This may be indicative of a relationship present at 16.4/11.2 ratios greater than 0.12, and 17.8/11.2 ratios greater than 0.03, but it is not conclusive. The integrated sources are shown to occupy three regions of the correlation plots: the data of \cite{boersma2010} at low abscissa values; the SINGS galaxies at an intermediate position; and \HII~regions or cirrus sources towards the upper-right quadrant. Together these produce a correlation (with coefficient $r=0.74$) but this is possibly spurious. Specifically, in the previous plots (Figure~\ref{fig:all_corrs1a}), the identified correlations can be seen both individually within our sample and individually in the sample of \cite{boersma2010}. In contrast, these correlations involving the 15.8 \mt band are not present in either sample in isolation. Additionally, the closeness of the [Ne~\textsc{iii}] 15.56 \mt line to the 15.8 \mt PAH band in low-resolution spectra affects how well the PAH band can be measured. This is compounded because the 15.8 \mt PAH band is weak and is near the inflection point of the rising 15-18 \mt continuum; the error on the continuum is not taken into account in the 3$\sigma$ cut-off. Finally, the Pearson correlation coefficient is known to be sensitive to outlying points \citep{devlin1975}, which is why the measured coefficients here are not necessarily meaningful, unlike in Figure~\ref{fig:all_corrs1a}.

\begin{figure}
	\includegraphics[width=\linewidth]{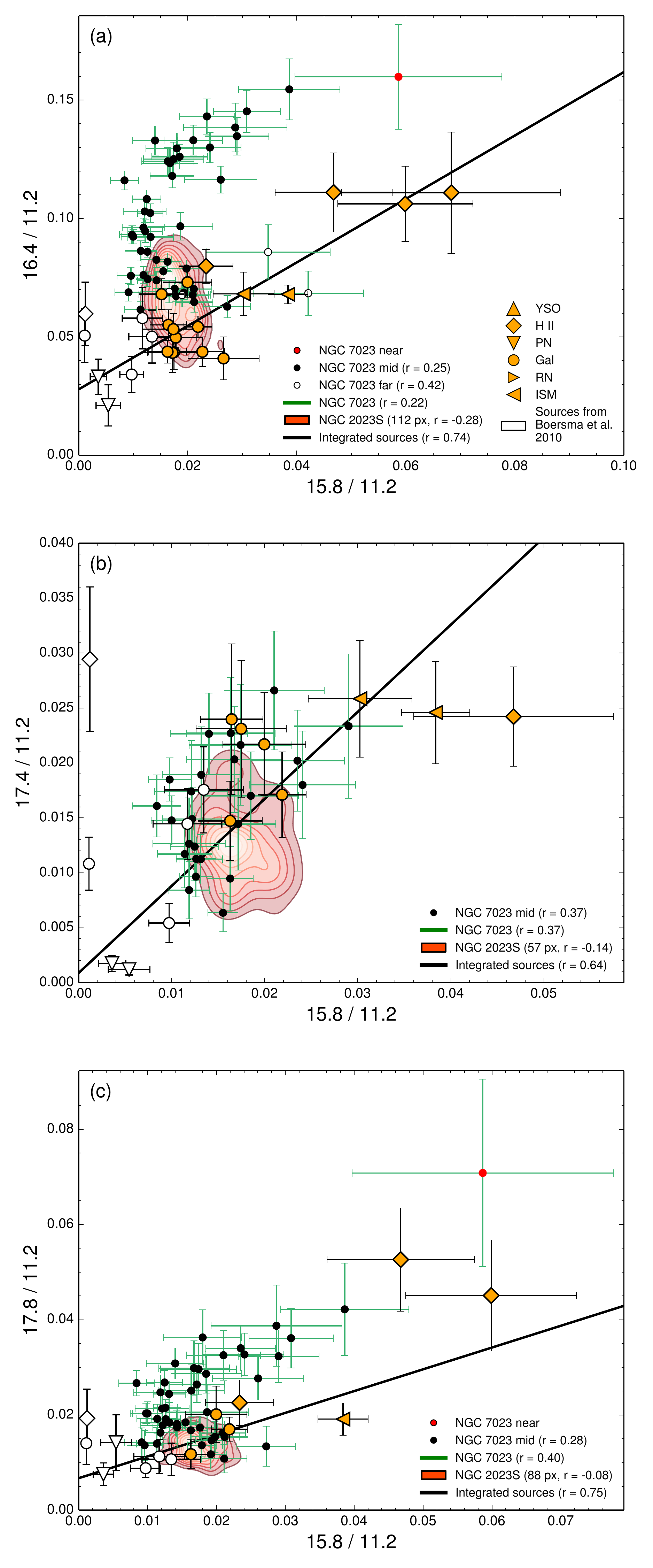}
	\caption{
	Correlation plots between the 11.2, 15.8 \mt bands and the 15-20 \mt plateau (panels a, b, c, respectively).
	}
	\label{fig:all_corrs1b}
\end{figure}

The 15.8 and 11.2 \mt bands were shown to correlate in NGC 2023 (paper \textsc{I}). We revisit this relationship in Figure~\ref{fig:all_corrs2_2}(a). The data near the peak PAH emission region (i.e., in the middle of the NGC 7023 map, see Figure~\ref{fig:ngc7023maps}) are shown to correlate slightly ($r=0.46$) and exhibit a gradient consistent with that in NGC 2023S. No correlation is identified for the integrated sources, though \HII~regions tend to lie at relatively high 15.8/12.7 ratios when compared to evolved stars or galaxies. The derived fluxes of the 15.8 and 12.7 \mt bands are influenced by the presence of the [Ne~\textsc{II}] 12.81 \mt and [Ne~\textsc{iii}] 15.56 \mt emission lines. These lines are present in \HII~regions and PNe and may influence or explain the observed discrepancy. Figures~\ref{fig:all_corrs2_2}(b) and (c) compare the emission of the 15-18 \mt plateau to those of the 11.2 and 15.8 \mt bands, respectively, which were shown to correlate in NGC 2023 (paper \textsc{I}). We find that NGC 7023 has similar correlations within its peak emission region; the other NGC 7023 regions do not show a clear correlation by themselves but they are consistent with the correlation seen in the NGC 7023 peak (``mid") and NGC 2023. The integrated sources show no linear trend but there is significant clustering near the position (Plat./12.7, 11.2/12.7) $\sim (1.5, 3.5)$.

\begin{figure}
	\includegraphics[width=\linewidth]{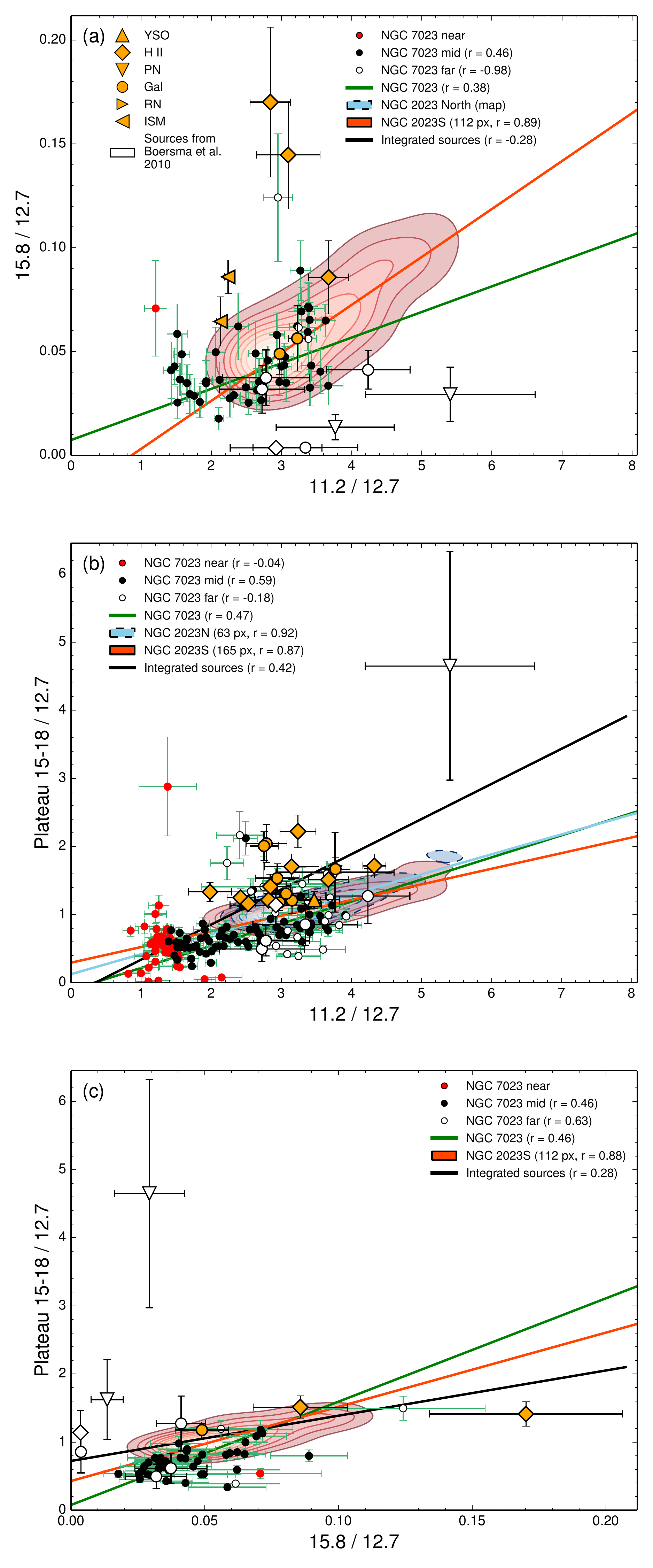}
	\caption{
	Correlation plots involving the 11.2 and 15.8 \mt bands and the 15-18 \mt plateau. (a) The 15.8 and 11.2 \mt bands, normalized to the 12.7 \mt PAH flux, correlate in RNe. (b), (c) 15-20 \mt plateau correlations with the 11.2 and 15.8 \mt bands in RNe.
	}
	\label{fig:all_corrs2_2}
\end{figure}

\subsection{Spatial distribution of 15-20 \mt PAH emission}
\label{sec:maps7023}

\subsubsection{Map morphology}

\begin{figure*}
	\centering
	\includegraphics[width=1.0\linewidth]{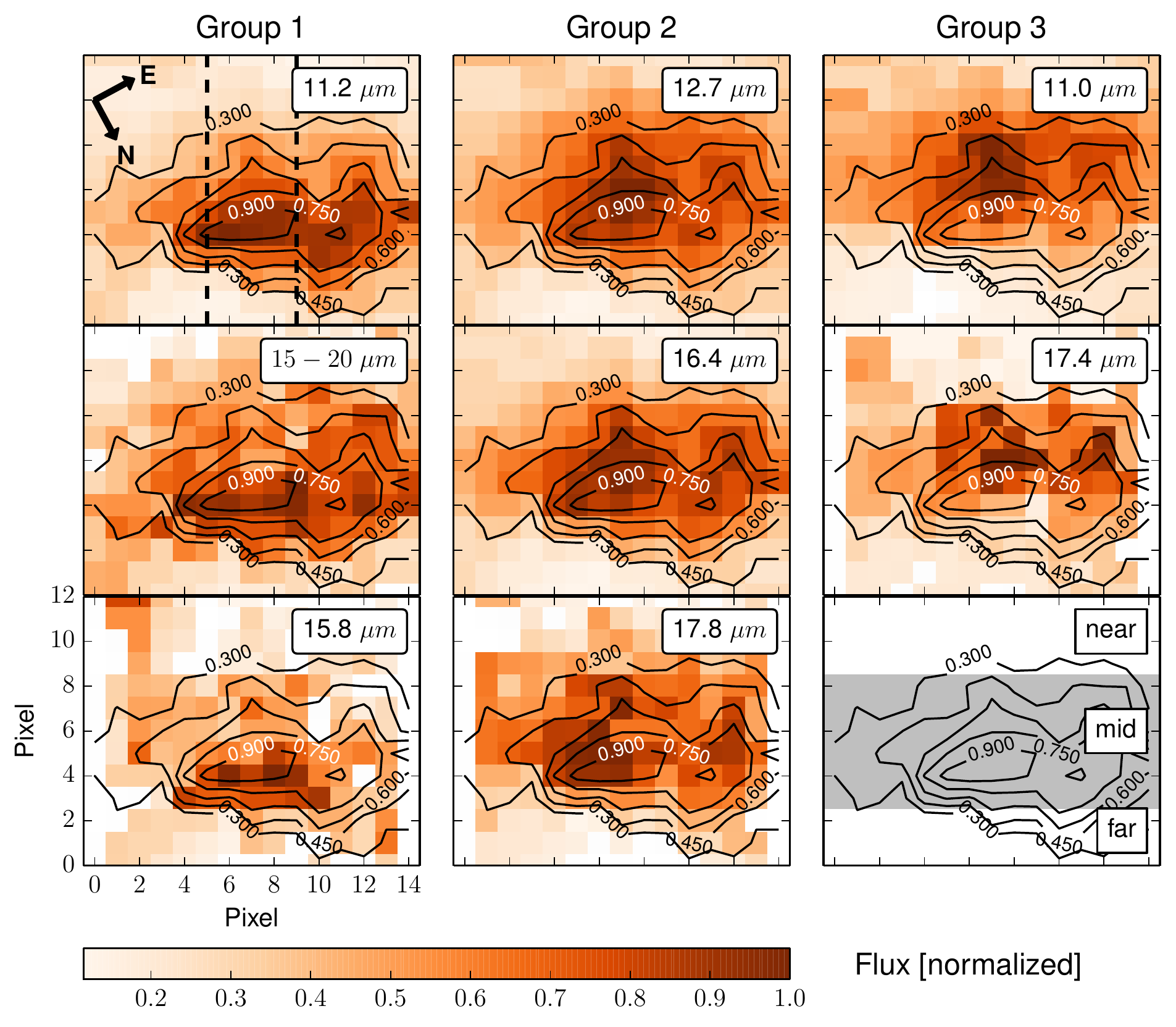}
	\caption{
	Maps of the normalized PAH band fluxes towards NGC 7023 ($13\times15$ pixels). Maps have been rotated 20$^{\circ}$ such that the star is at the top of each panel in this orientation. The original field of view is as indicated in Figure~\ref{fig:ngc7023}. The black contours are constructed from the 11.2 \mt band flux, normalized to unity. Bands are organized into vertical groupings (columns) based on similar characteristics (see text). The vertical dashed lines (shown only on the 11.2 \mt map) denote the region used for the 5 pixels-wide radial cut (Figure~\ref{fig:ngc7023slice}). We introduce in the lower-right panel three simple zones to distinguish regions of NGC 7023 data present in the correlation plots.
	}
	\label{fig:ngc7023maps}
\end{figure*}

We have thus far focused solely on correlations between bands. Paper \textsc{II} finds that the tightness of a correlation between any two bands is related to their spatial coincidence. However, even two bands with different spatial distributions could retain high correlation coefficients (e.g., the 8.6 and 7.7 \mt bands, with $r=0.936$; paper II). It is therefore important to determine if the correlations we identify are corroborated by their spatial morphologies. To do this, we examine the spectral map of NGC 7023 (Figure~\ref{fig:ngc7023maps}) and compare it with the spectral map of NGC 2023 (papers I and II). The 16.4 \mt data from this spectral cube has been previously examined by \cite{sellgren2007}, and the 11.0, 11.2, 12.7, 16.4 \mt bands and plateau emission by \cite{boersma2013,boersma2014,boersma2015}. Additionally, zones parallel to the PDR front have been identified by \cite{berne2012} and \cite{boersma2014}; we introduce three such zones in Figure~\ref{fig:ngc7023maps}.

The spatial distributions of the 12.7, 16.4 and 17.8 \mt bands are very similar in appearance. Each of these is seen to be displaced slightly toward the irradiating star relative to the 11.2 \mt band. The 17.8 \mt band is slightly more compact than the 12.7 and 16.4 \mt bands; we note it also has a limited SNR in comparison. In NGC 2023, the 12.7 and 16.4 \mt bands are also nearly identical in appearance, and the 17.8 \mt band is mildly more compact. The 17.4 \mt band has some apparent differences from the 12.7, 16.4 and 17.8 \mt bands in the spatial maps, despite correlating with these bands: its emission is more compact and appears to peak 10\arcsec~closer to the exciting star in both NGC 7023 and NGC 2023, though the 17.4 \mt band has a limited SNR in NGC 7023. The 11.0 \mt band is also displaced closer to the exciting star to a similar degree in both RNe. The 17.4 \mt band appears identical to the 11.0 \mt band in NGC 2023 prior to removing the C$_{60}$ component; after correction, the 17.4 \mt band is slightly more compact than the 11.0 \mt band.

The 11.2 \mt band and the 15-18 \mt plateau have coincident peak positions in both RNe, while the 15.8 \mt band peaks slightly further distant from the star. The 15.8 \mt band is much less extended than the 11.2 \mt band and the plateau in NGC 7023, but it is only reliably measured in a small portion of the map due to its low SNR. In NGC 2023, the 15.8 \mt band has a higher SNR and it shows a spatial distribution extremely close to that of the 11.2 \mt band. The 15-18 \mt plateau in NGC 2023 is much broader in extent.

\begin{figure}
	\centering
	\includegraphics[width=0.95\linewidth]{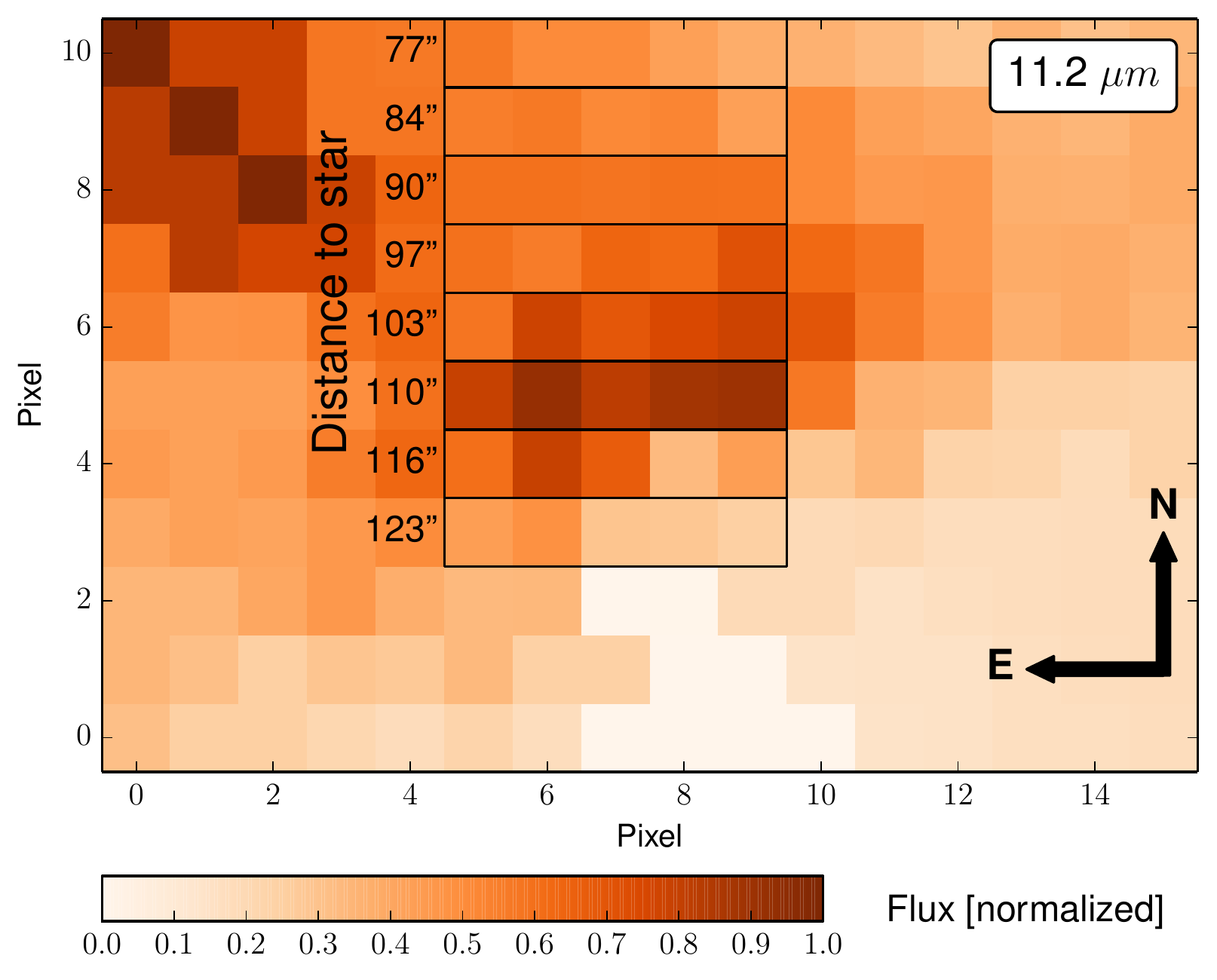}
	\caption{
	A normalized 11.2 \mt PAH intensity map of NGC 2023 S (papers \textsc{I} and \textsc{II}). These data have been rotated 35 degrees counter-clockwise to be roughly perpendicular to the irradiating star's radial vector. The outlined regions identify the 5 pixels-wide apertures used to create the radial slices. The exciting star is located upward in this projection, with estimated distance to each aperture as shown.
	}
	\label{fig:ngc2023map}
\end{figure}

\subsubsection{PAH emission variability with projected distance}

In order to investigate the change in PAH emission with projected distance from the illuminating star, we introduce radial slices or cuts across the maps of NGC 7023 and NGC 2023 (Figures~\ref{fig:ngc7023maps} and~\ref{fig:ngc2023map}) and analyze the individual band intensities as a function of the strength of the radiation field (Figures~\ref{fig:ngc7023slice} and ~\ref{fig:ngc2023slice}).

NGC 7023 is irradiated by HD 200775, a star of spectral type B2.5 Ve \citep{finkenzeller1985}, of effective temperature $17000K \pm 1000K$ \citep{baschek1982}. We find the strength of the FUV radiation field, G$_0$, to be $1.2\times10^5$ at 12\arcsec~and $1.4\times10^4$ at 35\arcsec~from HD 200775.\footnote{G$_0$ is the FUV flux between 6-13.6 eV, expressed in terms of the Habing field, $1.2\times10^{-4}$ erg/s/cm$^2$/sr \citep{habing1968}. We use the formulation of \cite{tielensbook}: $G_0 = 625 \frac{L \chi}{4 \pi d^2}$, where $L$ is the luminosity of the star, $\chi$ is the luminosity fraction between 6-13.6 eV and $d$ is the distance from the star. $\chi$ was determined by evaluating a blackbody of effective temperature corresponding to the appropriate stellar spectral type.} This is consistent with \cite{berne2012}, who adopted values of $1\pm0.7\times10^5$ at 12\arcsec~and $1\pm0.7\times10^4$ at 35\arcsec.

The slices of NGC 7023 show that the 11.2 and 15.8 \mt bands and the 15-18 \mt plateau peak furthest from the star, at a projected distance, d, of 49\arcsec, or G$_0\sim7.0\times10^3$. The 12.7, 16.4 and 17.8 \mt bands peak closer to the star, at d$\sim39$\arcsec$-44$\arcsec, or G$_0\sim11\times10^4-8.5\times10^3$. The 17.4 \mt band peaks at d$\sim40$\arcsec (G$_0\sim1.2\times10^4$) and the 11.0 \mt band has a wide peak from d$\sim30-39$\arcsec, corresponding to G$_0\sim1.9\times10^4-1.1\times10^4$. The 17.4 \mt band profile is much narrower than the 11.0 \mt emission, but it is aligned with the far edge of the 11.0 \mt band. Slices of the 16.4 and 18.9 \mt bands in NGC 7023 were presented by \cite{sellgren2007}. These authors observed the 16.4 \mt emission to peak at approximately 36\arcsec~distant from the star. We observe this peak to lie more distant, near 44\arcsec, but we use a different decomposition method. Overall, the spatial profiles of the 12.7, 16.4 and 17.8 \mt bands are broader and more symmetric than those of the other bands. We find that we can reproduce the profiles of the 12.7 and 16.4 \mt bands extremely well by adding the 11.0 and 11.2 \mt profiles and renormalizing the result (see Figure~\ref{fig:ngc7023slice}).

We perform a similar analysis on NGC 2023S, who's data have higher SNR and overall intensities than NGC 2023N's (Figure~\ref{fig:ngc2023slice}). NGC 2023 is illuminated by a B1.5V star \citep{mookerjea2009}. We derive G$_0$ values in a similar manner as for NGC 7023. The PAH transitions occur from G$_0\sim1.5\times10^3-3.0\times10^3$, in rough agreement with the estimates of $10^3 - 10^4$ in the literature, albeit on the low end (\citealt{sheffer2011}, paper \textsc{I}). The spatial stratification seen in NGC 7023 is also present in NGC 2023S. However, the transition occurs at quite different values of G$_0$ than NGC 7023. This suggests that our simple approximation of G$_0$ is not accurate enough and other parameters such as density, gas temperature and/or projection effects may play a role. Specifically, the 11.0 and 17.4 \mt bands peak nearest the star, with the 11.0 \mt band showing a broad peak from approximately 80-95\arcsec~distant (G$_0\sim2.9\times10^3-2.1\times10^3$), and the 17.4 \mt band being more compact near a distance of 84\arcsec (G$_0\sim2.6\times10^3$). The 11.2, 15.8 \mt bands and the 15-18 \mt plateau all peak near d$\sim$111\arcsec (G$_0\sim1.5\times10^3$), but we note the plateau is slightly broader. The 12.7 and 16.4 \mt bands are flat from d$\sim$80$-$105\arcsec~(G$_0\sim2.9\times10^3-1.7\times10^3$), again broader than the other emission bands, as in NGC 7023. The 17.8 \mt band is similar to the 12.7 and 16.4 \mt bands but it peaks closer to the star (80-95\arcsec, G$_0\sim2.9\times10^3-2.0\times10^3$), with a profile near that of the 11.0 \mt band. Again, the 12.7 and 16.4 \mt band profiles are well realized by summing and renormalizing the 11.0 and 11.2 \mt bands, albeit with a slight overestimate near 110\arcsec.

\begin{figure}
	\centering
	\includegraphics[width=1\linewidth]{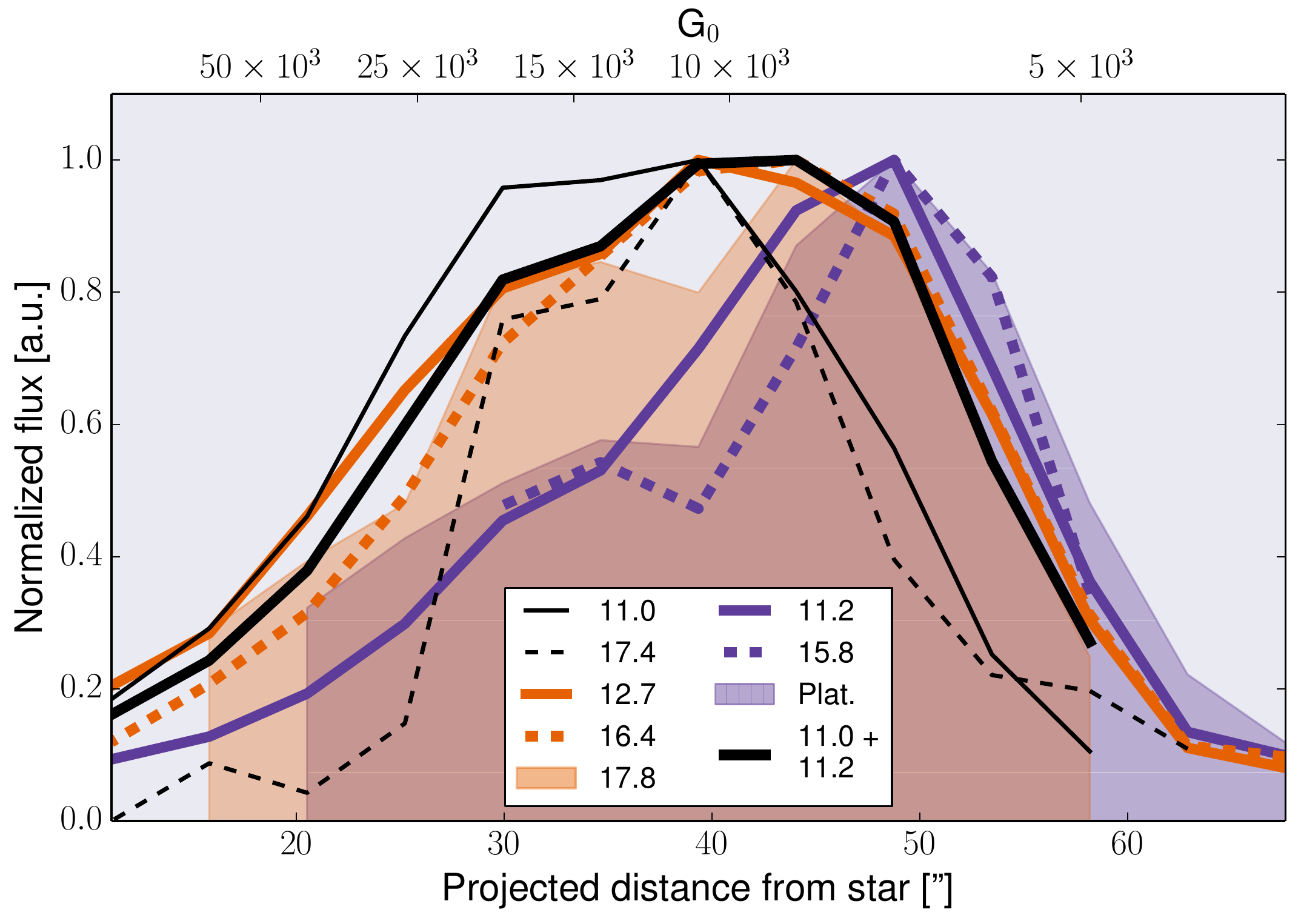}
	\caption{
Normalized PAH intensity along a projected cut across NGC 7023 directed towards the irradiating source, extracted at 12 positions, using a 5 pixels wide extraction aperture (see Figure~\ref{fig:ngc7023maps}). The colors indicate groupings of bands with similar peak positions and/or spatial profiles along the cut.
	}
	\label{fig:ngc7023slice}
\end{figure}

\begin{figure}
	\centering
	\includegraphics[width=1.0\linewidth]{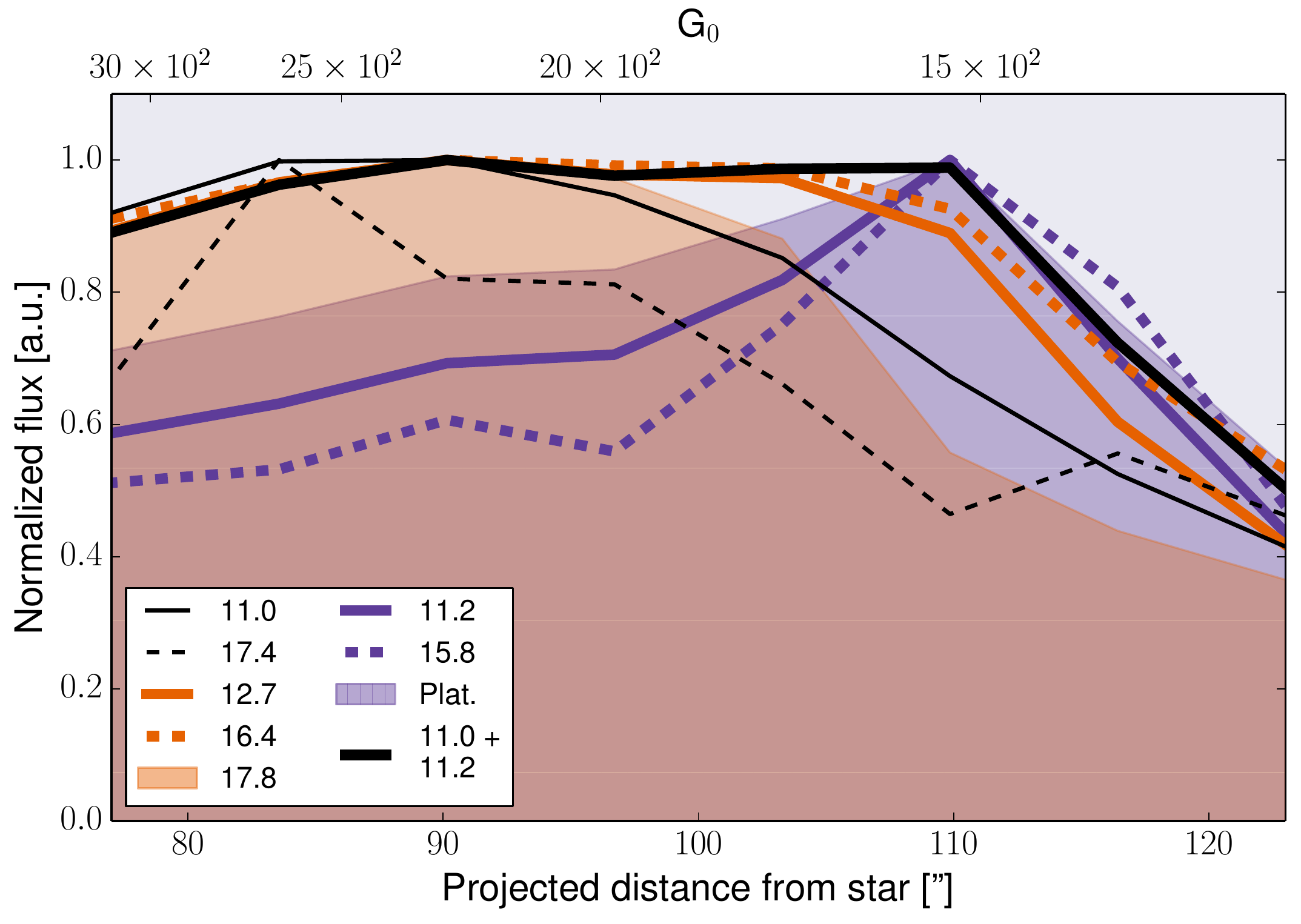}	
	\caption{
	Normalized PAH intensity along a projected cut across NGC 2023S directed towards the irradiating source, extracted at 8 positions, using a 5 pixels wide extraction aperture (see Figure~\ref{fig:ngc2023map}). The colors indicate groupings of bands with similar peak positions and/or spatial profiles along the cut.
	\vspace{0.1cm}
	}
	\label{fig:ngc2023slice}
\end{figure}

\section{Discussion}

\subsection{Interpretation}

We first summarize the observational results, and subsequently, we discuss the implications in terms of charge and molecular structure, based on comparison with theoretical studies. The observational results consist of PAH band intensity correlations for integrated and extended sources, as well as spatial measurements of PAH emission in spectral maps of NGC 7023 and NGC 2023.

\textit{Correlations.} The 16.4, 17.4 and 17.8 \mt band fluxes are inter-correlated in all environments. Furthermore, each of these three bands correlates with the 12.7 \mt band in the RNe. Amongst the sample of integrated sources, the 16.4 and 17.8 \mt bands correlate with the 12.7 \mt band only in \HII~regions. Together, these four bands (12.7, 16.4, 17.4 and 17.8 \m) correlate strongly with the 11.0 \mt band in RNe. The 15.8 \mt band does not correlate with any of these bands, leaving it as the odd one out. However, the 15.8 \mt band is inter-correlated in RNe with the 15-18 \mt plateau and the 11.2 \mt band. Finally, the 16.4 and 6.2 \mt bands show no overall linear correlation for the integrated sources, though if we exclude \HII~regions from this analysis they show a modest correlation, in line with results found in the literature \citep{boersma2010,boersma2014}. Our data are also isolated to relatively low ordinate and abscissa values, falling upon the lower end of the relationship found by \cite{boersma2014}.

\textit{Spatial maps.} The RNe spatial maps show a common morphology between the 12.7, 16.4 and 17.8 \mt bands. Similarity is also observed between the 11.2 and 15.8 \mt bands and the 15-18 \mt plateau. Lastly, the 11.0 and 17.4 \mt bands appear to peak in similar locations. We invoked radial cuts to isolate the spatial origin of each band to examine these similarities. The 11.0 and 17.4 \mt bands peak closest to the star, in similar locations, though the 11.0 \mt band's spatial profile is wider. The 11.2, 15.8 \mt bands and 15-18 \mt plateau peak furthest from the star, sharing very similar profiles. The 12.7 and 16.4 band profiles peak in an intermediate location, with very similar shapes. The 17.8 \mt band follows the 12.7 and 16.4 \mt bands in NGC 7023 but its profile is more similar to that of the 11.0 \mt band in NGC 2023S. We observe that the profiles of the 12.7 and 16.4 \mt bands bands (and the 17.8 \mt band in NGC 7023) can be reproduced by averaging the spatial profiles of the 11.0 and 11.2 \mt bands.

\subsubsection{Charge}

We first examine the observational results in the context of ionization states. The 11.0 and 11.2 \mt PAH bands have been firmly assigned to cationic and neutral PAHs, respectively \citep{hudgins1999,hony2001}. The similarities of the 15.8 \mt band to the 11.2 \mt band in the maps, cuts, and the fact that they correlate suggest that the 15.8 \mt band originates in neutral PAHs. Likewise, the observational data suggests the 17.4 \mt band is coincident with the 11.0 \mt emission, with which it correlates; we therefore associate the 17.4 \mt band with cationic PAHs. The 12.7, 16.4 and 17.8 \mt bands are self-similar in the maps and cuts, and have an extent that envelops both the 11.0 and 11.2 \mt bands. In conjunction with the correlation plots, these data suggest that the 12.7, 16.4 and 17.8 \mt bands arise from both cationic and neutral PAHs.

Paper \textsc{I} suggested that the 17.8 \mt band arises from both neutral and ionized PAHs, which we here confirm. Furthermore, paper \textsc{I} concluded that the 16.4 \mt band is from species cospatial with the cationic portion of the emitting PAH population. This result was strongly supported by the results of \cite{boersma2014}; the authors however noted that residual curvature in their correlations plots indicated that other population changes are at play. Here we identify two components of the 16.4 \mt band: one cospatial with cationic PAHs and one cospatial with neutral PAHs. The 12.7 \mt band was also shown to be coincident with PAH cations in paper \textsc{I}. \cite{rosenberg2011} used blind signal separation to identify two components of the 12.7 \mt band, which they associate with neutrals and cations. \cite{boersma2014} supported their assignment; this work also supports a combined neutral and cationic origin for the 12.7 \mt band. Paper \textsc{I} suggested the 17.4 \mt bands arises from doubly ionized PAHs and/or a subset of the cationic PAH population. We attribute the 17.4 \mt bands to cations generally, and not necessarily a particular subset. The discrepancy may be due to how each paper traces cations: we use the 11.0 \mt band as a tracer of cations here, whereas paper \textsc{I} used the 6.2 and 7.7 \mt bands as tracers. Paper \textsc{II} finds that there are two populations contributing to the 7.7 \mt band (and thus the 6.2 \mt band, due to its similar morphology), and so the 7.7 \mt band may not be an ideal tracer of PAH cations. Additionally, \cite{boersma2015} show that the 7.7 \mt PAH band strength may not be a reliable tracer for PAH cations. We attribute the 15.8 \mt band to neutral PAHs, in agreement with paper \textsc{I}. 

\vspace{5mm}

Theoretical studies of PAH vibrational modes have given us clues about the origins of the 15-20 \mt emission bands \citep{boersma2010,ricca2010}. These often make use of the NASA Ames PAH IR Spectroscopic Database\footnote{\url{http://www.astrochem.org/pahdb/}} \citep{bauschlicher2010,boersma2014_amesdb}. Specifically, the 16.4 \mt band appears to arise from a mixture of out-of-plane and in-plane vibrations, particularly due to elongation and compression modes \citep{ricca2010}. The 17.4 \mt band originates in a mixture of C-C-C in-plane and out-of-plane modes and the 17.8 \mt band is attributed to a mixture of C-H and C-C-C out-of-plane bending. The origin of the 15.8 \mt band is unclear, though emission at 15.4 \mt by C-C-C in-plane bending modes was identified by \cite{ricca2010}; this may or may not correspond to the astronomical 15.8 \mt band. Using these analyses, we assert that the intrinsic intensities of the 15.8, 16.4, 17.4 or 17.8 \mt bands do not vary by more than approximately a factor of 2 between charge states (anions, neutrals, cations). This estimate is based on Figure 13 of \cite{ricca2010}, which presents average spectra for all PAHs in the Ames database containing more than 50 carbon atoms, separated by ionization state, and Figure 15 of \cite{boersma2010}, showing the intrinsic intensities for different charge states of C$_{96}$H$_{24}$. In general then, the 15-20 \mt emission bands produce intensity variations that are much more subdued than the variations observed in the 3-12 \mt range, which span approximately one order of magnitude \citep{allamandola1999}. Emission by anions, neutrals and cations is expected for all four 15-20 \mt bands in the database \citep{ricca2010}. The 16.4 \mt band is expected to be twice as strong for cations than for the other charge states \citep{bauschlicher2010,ricca2010}. The 17.8 \mt band appears to have similar intensities in each charge state, while the 17.4 \mt band should be strongest in anions by a factor of two \citep{ricca2010}. These authors computed spectra for eight neutral PAHs and noted similar intensities between the 17.4 and 17.8 \mt band, possibly implying a correlation; such a correlation is observed in this work. Lastly, emission was observed at 15.4 \mt by \citep{ricca2010}, with cations exceeding the intensities of neutrals and anions by a factor of two, but it is unknown if this emission corresponds to the astronomical band at 15.8 \m.

\vspace{5mm}

Setting aside anions due to their expected very low abundances, we note that the prediction of similar emission intensities of the 17.8 \mt band between cations and neutral PAHs is consistent with our observational results. This can be seen in the spatial maps and cuts, in which the 17.8 \mt band's spatial profile encompasses the 11.0 and 11.2 \mt bands, centered between them. The 16.4 \mt band, which is predicted to be twice as strong in cations than neutrals, is however somewhat at odds with the observational data: we observe it has the same spatial profile as the 17.8 \mt band, with no preference or tendency towards the 11.0 \mt emission over the 11.2 \mt emission. This suggests it should arise from a similarly balanced mixture of cations and neutrals from which the 17.8 \mt band originates. The 17.4 \mt band is predicted by the database to have similar emission intensities between cations and neutral PAHs, which is inconsistent with our assignment solely to cations. Lastly, the database 15.4 \mt band is seen to be cation-dominant, at odds with the neutral character of the 15.8 \mt astronomical band from the observational data. It may simply be the case that these two bands do not correspond to the same emission feature. Other possibilities for the observed inconsistencies may include variable PAH population abundances, incompleteness and/or poor representation of these particular (sub-)populations in the database, or simply that our data are not sufficiently sensitive to probe variations at the scale of two-to-one. An examination of the radial cuts shows the latter point to be true.

\subsubsection{Molecular structure}

Molecular structure determines which vibrational modes are available to any particular PAH, but it is not known a priori how much of an influence structure will play in driving band intensity correlations on the whole. Some headway has been made in this area, in so far as particular band attributions have been suggested. The 16.4 \mt band has been attributed to PAHs containing pendent rings or ``pointed" edges (e.g., \citealt{vankerckhoven2000,peeters2004b,boersma2010,ricca2010}). The former are single cycles that attach to the carbon skeleton through two shared carbon atoms, while the latter are corners that naturally arise in parallelogram-shaped PAHs, attached through three shared carbon atoms. The 16.4 \mt band correlates well with the 17.8 \mt band, which appears to be somewhat structure insensitive: the 17.8 \mt band was present in all eight large neutral PAHs studied by \cite{ricca2010}, despite encompassing a variety of structures, including some with pointed edges. The 12.7 \mt band, which correlates well with both the 16.4 and 17.8 \mt emission, is associated with a combination of duo and trio C-H out-of-plane modes \citep{hony2001,bauschlicher2009,bauschlicher2010}. Recently, this band was suggested to specifically originate in PAHs having an armchair-like structure \citep{candian2014}. As pendent rings and pointed edges naturally constitute duos and trios, it is then possible that the 12.7 and 16.4 \mt bands naturally correlate due to these common features. The 17.8 \mt band is not known to have a dependence on duos or trios, but it has been suggested that it is a molecule-independent feature \citep{ricca2010}, which may explain why it correlates with the 12.7 and 16.4 \mt bands.

The 17.4 \mt band, which correlates with the 12.7, 16.4 and 17.8 \mt bands, generally appears to show the strongest emission intensities in compact PAHs \citep{bauschlicher2010,ricca2010}. The 15.6 \mt band (which may or may not correspond to the astronomical 15.8 \mt band) has been suggested to arise from vibrations of pendent rings \citep{boersma2010}. Pendent rings are also suggested as one possible origin for the 16.4 \mt band. If both the 15.8 and 16.4 \mt bands are (at least in part) due to pendent rings, one would expect an intensity correlation, depending on environmental conditions (which we cannot speculate on here). In general the observational data show no such correlation between these bands, though there is a slight hint of a linear correlation when the ratio of 16.4/11.2 exceeds $0.12$ (see Figure~\ref{fig:all_corrs1b}). The correlation of the 15.8 \mt band with the 11.2 \mt band is suggestive of large molecules, but it is known that the emission intensity of the 15.8 \mt band decreases with increasing PAH size in the NASA Ames PAH IR Spectroscopic Database \citep{bauschlicher2010,ricca2010,boersma2014_amesdb}. Finally, \cite{boersma2010} examined classes of PAHs, including compact, irregular, and pentagon-containing PAHs. Minor systematic trends were observed, including the fact that irregular PAHs tend to have richer and more variable spectral features when compared to compact PAHs. As molecule size increased, compact PAHs exhibited little variation in the types of C-C-C vibrational modes that were active, whereas the complex edge structure of irregular PAHs facilitated a larger variety of C-C-C vibrational modes \citep{boersma2010}. Even with this knowledge, at the present there is no clear mapping between molecular structure and the observational data.

\subsubsection{Summary}

The exact balance between charge and molecular structure in driving the observed correlations of the 15-20 \mt bands cannot be discerned. However, the maps and radial cuts appear to cleanly separate these bands into those dominated by cations (11.0, 17.4 \m), neutrals (11.2, 15.8 \m, 15-18 \mt plateau), or a combination of the two (e.g., 12.7, 16.4, 17.8 \mt bands). This suggests charge is the dominant factor driving these correlations, though there are indications that some bands should correlate due to structural arguments (e.g., the 12.7 and 16.4 \mt bands). Ultimately, theory points towards a mixture of highly variable emission features (e.g., \citealt{boersma2010}, their Figures 11-15), whereas in space we observe a relatively simple picture of four bands and one underlying plateau. This simplicity may reflect a basic commonality of the carriers, or suggest that a relatively small collection of stable molecules are responsible for the observed emission bands \citep{boersma2010}. The latter is known as the \textit{grandPAH} hypothesis \citep{andrews2015}.

\subsection{Statistics of the 15-20 \mt emission}
\label{sec:stats}

First we determine if the relative emission strengths of the 5-15 \mt bands in our sample are consistent with similar measurements in the literature \citep{peeters2002}. We accomplish this by measuring the fractional contribution of the 6.2, 7.7, 8.6 and 11.2 \mt bands to the total MIR emission (which we define here as the total flux emitted by these bands) for nine \HII~regions and fifteen galaxies in our sample. Since \cite{peeters2002} only had two galaxies in their sample, we use the \HII~regions as a direct point of comparison. Our results are presented in Figure~\ref{fig:statistics}, panel (a). We find that the \HII~regions are consistent with those of \cite{peeters2002}, whose data is presented in panel (b), and there is no discrepancy between \HII~regions and galaxies in our sample. Note that \cite{peeters2002} included the 3.3 \mt band and not the 8.6 \mt band in their analysis, whereas we do the opposite (as our spectra start at 5 \m). The 3.3 and 8.6 \mt bands are quite weak (less than 10\% of the total MIR emission in both studies) and should therefore not strongly influence the analysis.

The same analysis is performed for the ``FIR" features, i.e., the 15.8, 16.4, 17.4, 17.8 \mt bands and the 15-18 \mt plateau, shown in Figure~\ref{fig:statistics}, panel (c). Due to the inconsistent detections of the 15.8, 17.4 and 17.8 \mt bands, the 16.4 \mt band and the 15-18 \mt plateau are used together as a proxy for the total FIR PAH flux. The 15.8, 16.4 and 17.8 \mt bands have fractional strengths in \HII~regions of $7\pm4$\%, $16\pm2$\% and $6\pm1$\%, respectively. In galaxies, these are systematically lower, with values $4\pm1$\%, $11\pm3$\% and $3\pm1$\%, respectively. The 17.4 \mt band is observed to have similar strengths between \HII~regions and galaxies ($3\pm1$\% in each). The plateau makes up the remainder of the measured fractional strengths. The observed inconsistency by object type may be due to underestimating the 17.4 \mt PAH emission in \HII~regions or overestimating the 17.4 \mt emission in galaxies. Neither would be surprising, as C$_{60}$ emission at 17.4 \mt is common, and it can only be removed by measuring the 18.9 \mt C$_{60}$ emission, which itself is blended with the [S~\textsc{iii}] 18.71 \mt line. Recall the typical 17.4/18.9 \mt C$_{60}$ band strength ratio is approximately 0.5 \citep{cami2010,sellgren2010,bernardsalas2012}.
Another possibility is that the 15.8, 16.4 and 17.8 \mt bands are all overestimated in \HII~regions, which requires a trio of systematic errors that we deem unlikely, as the same method was applied to both \HII~regions and galaxies. The third possibility is that these bands truly are systematically higher in \HII~regions than galaxies.

As a separate analysis, we investigate whether the MIR/FIR ratio depends on object type, where the 11.2 and 16.4 \mt bands are used as proxies for the MIR and FIR emission, respectively. We therefore examined 490 low-resolution spectra of the SAGE-Spec sample and used the following selection criteria: 1. PAHs must be clearly present; 2. the spectra displayed no discontinuous jumps; 3. sources were less extended than 9\arcsec (which corresponds to the approximate size of the SL aperture at 6.2 \m); and 4. the 11.2 and 16.4 \mt PAH bands must have 3$\sigma$ detections. We identify 10 \HII~regions and 22 evolved stars as meeting these criteria. The \HII~regions have a mean 16.4/11.2 ratio of $0.09 \pm 0.02$. Evolved stars exhibited a ratio of $0.12 \pm 0.05$, consistent with the \HII~regions. As such, we find no dependence on object type for the relative MIR and FIR band intensities in this analysis.

\begin{figure}
	\includegraphics[width=\linewidth]{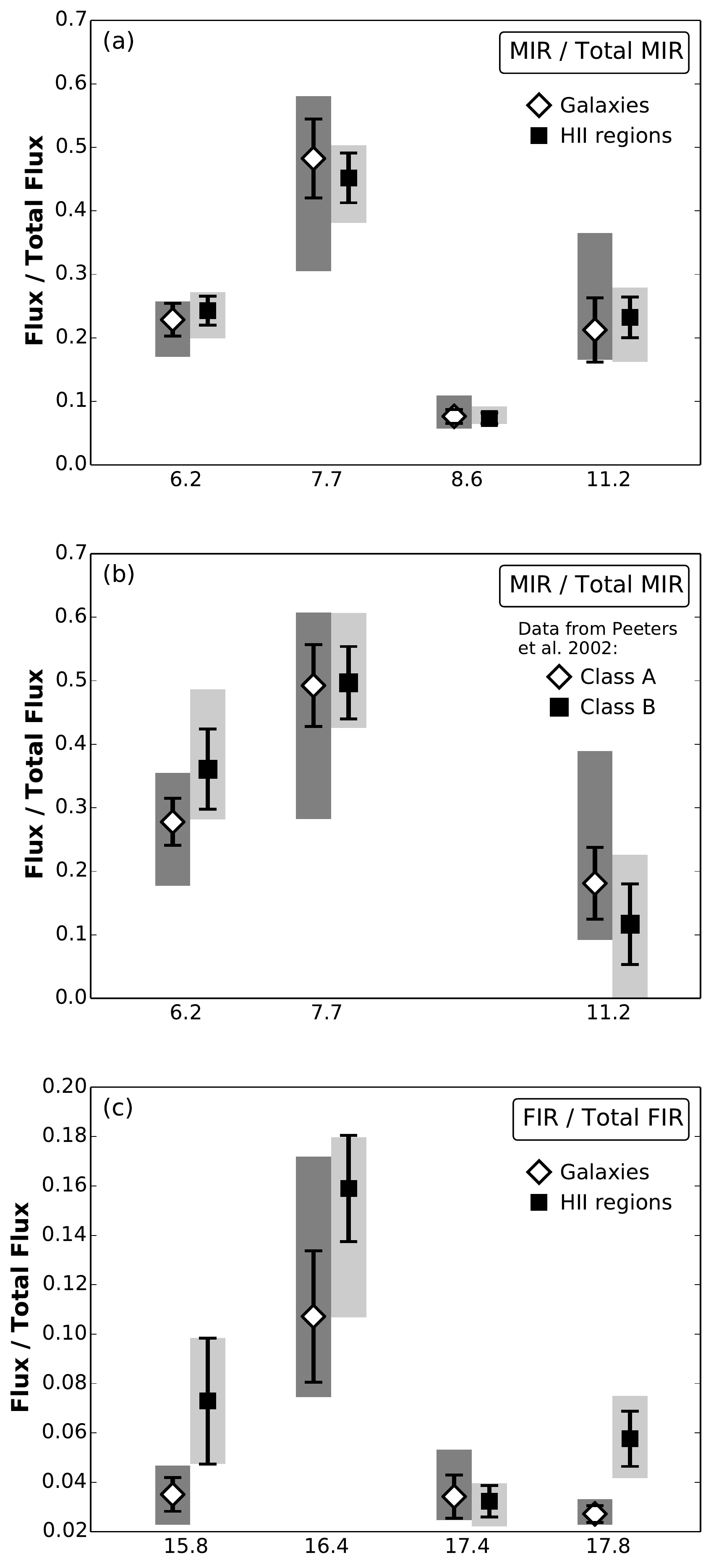}
	\caption{
	Statistical summary of the 5-15 \mt and 15-20 \mt PAH bands. The fraction of total flux in each band is illustrated by its mean (white diamond for galaxies, black square for \HII~regions), standard deviation (black line), and minimum and maximum values (denoted by the grey rectangles). (a) The 6.2, 7.7, 8.6 and 11.2 \mt emission (called ``MIR" here) are normalized to their combined total emission. (b) The same analysis, for class A and B sources, as presented in \cite{peeters2002}. Note the authors included 3.3 \mt emission, but did not have 8.6 \mt emission. (c) The fractional emission in the ``FIR" bands (15.8, 16.4, 17.4, 17.8 \mt bands) are compared. The total FIR emission is considered to be the sum of the 16.4 \mt and plateau emission. The plateau fraction ($\gtrsim0.7$) is omitted for clarity. 
	}
	\label{fig:statistics}
\end{figure}

\section{Summary \& Conclusions}

The 15-20 \mt PAH emission bands have been investigated in fifty-seven sources, comprised of LMC point sources from the SAGE-Spectroscopy survey, nearby galaxies from SINGS, two extended cirrus sources and a spectral map of NGC 7023. We also included the sample of \cite{boersma2010} and spectral maps of NGC 2023 (papers \textsc{I} and \textsc{II}). We investigated correlation plots of band flux ratios and examined the map morphologies of the RNe NGC 7023 and NGC 2023. We performed radial cuts across these maps to evaluate the spatial profiles of the 15-20 \mt PAH emission. Our primary conclusions are as follows:

\begin{enumerate}
\item Correlation plots:

\begin{itemize}
\item
The 16.4, 17.4 and the 17.8 \mt bands are inter-correlated in all environments.

\item
Within the RNe we see that the 16.4, 17.4 and 17.8 \mt bands further correlate with the 11.0 and 12.7 \mt bands. In \HII~regions, the 12.7 \mt band also correlates with the 16.4 and 17.8 \mt bands.

\item
The 11.2 and 15.8 \mt bands and the 15-18 \mt plateau are inter-correlated in NGC 7023, consistent with results for NGC 2023 (paper \textsc{I}). The 15.8 \mt band does not correlate with the 11.0, 12.7, 16.4, 17.4 or 17.8 \mt bands in general. Within a sub-region of NGC 7023 the 15.8 \mt band appears to show a modest correlation with the 16.4 and 17.8 \mt bands.
\end{itemize}

\item Spectral maps:

\begin{itemize}
\item
The maps show similar morphologies between the 11.0 and 17.4 \mt bands, the 12.7, 16.4 and 17.8 \mt bands, and the 11.2, 15.8 \mt bands with the 15-18 \mt plateau.

\item
The radial slices in the RNe maps show that the spatial profiles of the 12.7, 16.4 and 17.8 \mt bands can be reconstructed by averaging the 11.0 and the 11.2 \mt band profiles.
\end{itemize}

\item
Charge dominates: we attribute the 12.7, 16.4 and 17.8 \mt bands to a combination of neutral and cationic molecules. The 17.4 \mt band we attribute to cations, based on similarity of its spatial distribution to the 11.0 \mt emission in the maps and radial cuts. Our results support the association of the 15.8 \mt band with PAHs, specifically neutral PAHs (paper \textsc{I}).

\item
The 16.4/11.2 ratio in the general SAGE-Spec sample shows no dependence on object type. The fractional emission of the 15.8, 16.4 and 17.8 \mt bands in \HII~regions are systematically higher than those in galaxies. The 17.4 \mt band is the exception, which has similar fractional strengths in both environments.

\end{enumerate}

With the significant number of sources in our sample, we were able to discover correlations between the 15-20 \mt features that were not discernible in the much smaller sample  studied by \cite{boersma2010}. The forthcoming JWST instrument MIRI will be an essential tool in furthering this analysis. Its large collecting area and increased spatial sensitivity will permit the construction of larger sample sizes, permitting further statistical analysis. Additionally, its improved spectral sensitivity will also allow us to examine the PAH band profiles in detail, with less line-blending than is possible with current data sets. These key improvements, in conjunction with theoretical and laboratory analysis, may help disentangle the inter-related behaviours of the 15-20 \mt bands.

\section*{Acknolwedgements}

We thank the anonymous referee for the useful feedback. The authors thank J.D. Smith for supplying the SINGS data, as well as C. Boersma for supplying the data of their study. The authors also thank P.M. Woods for providing object classifications. The authors acknowledge support from NSERC discovery grant, NSERC acceleration grant and ERA. The IRS was a collaborative venture between Cornell University and Ball Aerospace Corporation funded by NASA through the Jet Propulsion Laboratory and Ames Research Center \citep{houck2004}. This research has made use of NASA's Astrophysics Data System Bibliographic Services, and the SIMBAD database, operated at CDS, Strasbourg, France.

\bibliographystyle{apj}

\end{document}